\documentclass{aa}  

\usepackage{graphicx}
\usepackage{txfonts}
\usepackage{natbib}

\usepackage[english]{babel}
\usepackage{caption}
\usepackage{subcaption}
\usepackage{amssymb}
\usepackage{ucs}
\usepackage{amsmath}
\usepackage{amsfonts}
\usepackage{color}

\newcommand{\dd}{\mathrm{d}}
\newcommand*\rfrac[2]{{}^{#1}\!/_{#2}}

\newcommand{\corr}[1]{#1}

\newcommand{\LEt}[1]{}

\begin{document}

   \title{Variation of bulk Lorentz factor in AGN jets due to Compton rocket in a complex photon field}

   \author{T. Vuillaume
          \inst{1}
          \and
          G. Henri \inst{1}\fnmsep
          \and
          P-O. Petrucci \inst{1}
          }

   \institute{Institut de Plan\'etologie et d'Astrophysique de Grenoble,
              414 rue de la Piscine, St-Martin d'H\`eres, France\\
              \email{thomas.vuillaume@obs.ujf-grenoble.fr}
             }

   \date{}

 
  \abstract{   Radio-loud active galactic nuclei are among the most powerful objects in the universe. In these objects, most of the emission comes from  relativistic jets getting their power from the accretion of matter onto \LEt{ in? can matter fall on a hole?  }  supermassive black holes. However, despite the number of studies, a jet's acceleration to relativistic speeds is still poorly understood. \\
It is widely known that jets contain relativistic particles that emit radiation through several physical processes, one of them being the inverse Compton scattering of photons coming from external sources. In the case of a plasma composed of electrons and positrons continuously heated by the turbulence, inverse Compton scattering can lead to relativistic bulk motions through the Compton rocket effect. We investigate this process and compute the resulting bulk Lorentz factor in the complex photon field of an AGN composed of several external photon sources.\\
We consider various sources here: \LEt{ i.e., the inner parts of the photon field?   } the accretion disk, the dusty torus, and the broad line region. We take  their geometry and anisotropy carefully  into account in order to numerically compute the bulk Lorentz factor of the jet at every altitude.\\
The study, made for a broad range of parameters, shows interesting and unexpected behaviors of the bulk Lorentz factor,  exhibiting acceleration and deceleration zones in the jet. We investigate the patterns of the bulk Lorentz factor along the jet depending on the source sizes and on the observation angle and we finally show that these patterns can induce variability in the AGN emission with timescales going from hours to months.
}

   \keywords{Galaxies: jets --
                Galaxies: active --
                Radiation mechanisms: non-thermal --
                Relativistic processes --
                scattering --
                plasmas
               }

   \maketitle
%

\section{Introduction}


It is now widely known that AGN jets hold relativistic flows. The first evidence of this goes back to the 1960s with the  interpretation of brightness temperatures of quasar radio emission exceeding the Compton limit by \cite{1966Natur.211..468R}. This was beautifully confirmed by the observation of superluminal motions with the achievement of the very long base interferometry (VLBI) \LEt{ acronym to be introduced at first use}technique \citep{1971ApJ...170..207C}. These superluminal events are only possible for actual speeds larger than $0.7c$. Relativistic velocities are also required to avoid strong $\gamma-\gamma$ absorption by pair production, the high-energy photons being able to escape thanks to beamed radiation (see \citealt{1994ApJS...90..899B}). 

Relativistic flows are characterized by their bulk Lorentz factor $\Gamma_b=\left( 1- \beta_b^2\right)^{-1}$ rather than their speed $V_b$ with $\beta_b=V_b/c$. However, there are several pieces of evidence that this bulk Lorentz factor is not homogeneous throughout the flow. The spatial distribution of the relativistic motion in the jet is still a matter of discussion. Two types of variations are possible: radial and longitudinal (or a combination of both).
The variations of longitudinal bulk Lorentz factor have  often been parametrized by power laws with an accelerating and/or a decelerating phase (\citealt{Marscher:1980js}, \citealt{Ghisellini:1985wr}, \citealt{Georganopoulos:1998jf}, \citealt{Li:2004bc}, \citealt{Boutelier:2008bga}). Even though an initial accelerating phase appears necessary to achieve\LEt{ achieve ? } relativistic speeds, decelerating flows  have also been invoked at larger scales, for example, to unify BL Lacs and radiogalaxies. In the unification scheme, BL Lacs and FR I radio galaxies are the same type of objects seen at different angles. However, BL Lacs models with constant jet velocities need very high bulk Lorentz factors to produce the observed SED, which is in contradiction with the FR I models and the observed jet velocities at subparsec scales in the TeV BLacs Mrk 421 and Mrk 501 (Marscher 1999). In their work, \cite{2003ApJ...594L..27G} showed that deceleration of the jet allows photons emitted in the inner parts of it to be scattered by the upper parts. In this case, radiation from fast regions of the jet would be highly beamed and thus correspond to BL Lacs objects, while radiation from slower regions would be emitted in a wider cone and would correspond to radio galaxies. Implications of a bulk velocity structure for the observed spectral energy distribution (SED) has been studied by \cite{2009PASJ...61.1153Y}.

In addition, a radial distribution of velocity is possible and has been particularly studied in the case of a double-jet structure, the so-called spine/layer jet. Here too, the idea was proposed as a solution to the unification issue between BL Lac objects and radiogalaxies (see \citealt{2000A&A...358..104C}). In this framework, \cite{Ghisellini:2005bc} explain the rapid variability of the TeV emission without requiring huge Doppler factors, and \cite{2014arXiv1404.6894T} are able to reproduce the SED of NGC 1275. Recent observations (\cite{Giovannini:1999bja}, \citealt{Swain:1998bf}, \citealt{2004ApJ...600..127G})  also bring evidence of such structures.

The idea of a two-flow structure was first proposed by \cite{1989MNRAS.237..411S} for theoretical reasons. \LEt{ have I interpreted correctly? } In this paradigm, the jet is assumed to be made of two components: a mildly relativistic sheath composed of $e^- / p^+$ and driven by magnetohydrodynamical (MHD) forces\LEt{ introduce acronym}, which  transports most of the kinetic energy, and an ultra-relativistic spine composed of $e^- / e^+$ pairs, which is  responsible for most of the emission \LEt{ ok? I had to rearrange the punctuation. }. A detailed description of the formation of such a pair beam has been developed in \cite{1995MNRAS.277..681M} and following works (e.g., \cite{1998MNRAS.300.1047R}, \cite{2006ApJ...640..185H}, \cite{Boutelier:2008bga}). In this model, the outer jet acts as an energy reservoir for the particles of the spine. Starting from an initial injection of some relativistic particles (possibly created in the surrounding of a rotating black hole), these particles will emit copious amounts of high-energy radiation, which will be converted into pairs. These pairs will be in turn continuously reaccelerated along the jet via the second-order Fermi process through the turbulence triggered by various instabilities in the outer MHD jet. Observations of diffuse X-ray emission in type 1 Fanaroff-Riley objetcs \LEt{ used only here so should be spelled out }are in favor of this view of distributed particle acceleration rather than localized shocks (\citealt{2007ApJ...670L..81H}).

In leptonic models, X-ray and gamma-ray emission is thought to arise from inverse Compton (IC) process on soft photons. These photons can be provided by synchrotron emission (called the synchrotron self-Compton or SSC process), or by external sources such as an accretion disk, the broad line region, or a dusty torus. All these sources will give a locally anisotropic photon field on the axis of the jet. Under these conditions, the emitted radiation will  also be highly anisotropic, which produces a strong momentum transfer between the relativistic plasma and the emitted radiation, the so-called Compton rocket effect (see \citealt{1981ApJ...243L.147O}, \citealt{Melia:1989ep} or more recently \citealt{1998MNRAS.300.1047R}). 

\corr{
For a relativistic  $e^- / e^+$ pair plasma, this force is dominant and will drive the bulk motion to relativistic velocities. 
As is detailed in section \ref{sec:Geq}, this effect will  saturate when the velocity (or equivalently the bulk Lorentz factor) of the plasma reaches a characteristic value for which the net radiation flux vanishes in the comoving frame as a result of the aberration of the photon momentum. Thus, a plasma submitted to this radiation force will tend to reach this equilibrium velocity, which can be viewed as the average velocity of the photon ``wind''.
The Compton rocket effect can also be found under the name of Compton drag effect. Even though they are exactly the same mechanism, the drag denomination comes from an a priori assumption of a very high bulk Lorentz factor, higher than that imposed by the external radiation and the Compton rocket effect: in this case, the inverse Compton rocket (more appropriately called a ``retrorocket'' effect in this case)  will result in a deceleration of the flow or a limitation of its velocity (\cite{Sikora:1996uf} and \cite{2010MNRAS.409L..79G}).

The Compton rocket effect has often been  dismissed as a cause of relativistic motion because it is also a cooling process. This means that an isolated relativistic pair plasma will  also be quickly cooled and will  generally be unable to reach the high bulk Lorentz factors $(\approx 10)$ needed to explain superluminal motions (\citealt{Phinney:1982wt} and also \cite{2000ApJ...534..239M}). However, this objection is not valid \LEt{ is not valid? }in the two-flow model, since the relativistic pair plasma is supposed to be continuously reheated by the surrounding MHD flow.

Under these conditions, a pure electron-positron pair plasma can be coupled to the radiation field over a much larger distance. It will then stick to the  equilibrium velocity (which is generally variable),  until the radiation field weakens enough for the decoupling to occur. The plasma will then essentially follow a ballistic motion at the terminal velocity, which depends on the location of the decoupling. In the following, we will only study the value of the equilibrium velocity, which depends only on the radiation field and not on the characteristics of the plasma in the Thomson regime. On the other hand, the location of the decoupling, and hence the terminal velocity, depends on these plasma characteristics,  as is discussed in section \ref{sec:Ginf}.

In summary, we investigate this paradigm and propose to study the evolution of the resulting bulk Lorentz factor (presented in section \ref{sec:Geq}) due to the Compton rocket effect in a complex photon field including three main external sources of soft photons present in quasars,  the accretion disk, the dusty torus, and the broad line region (BLR) (presented in section \ref{sec:modeling}). By computing accurately the equilibrium bulk Lorentz factor along the jet in the Thomson regime (sections \ref{sec:Geq_res} and \ref{sec:param_evol_gam}), we can study what   effect it might have on the observed emission (sections \ref{sec:param_evol_gam} and \ref{sec:emission_var}).

In the rest of the paper, the ``jet'' refers to the inner spine of the two-flow model which is subject to the Compton rocket effect and is the flow moving at relativistic bulk speeds. Primed quantities are expressed in the comoving frame and unprimed quantities are expressed in the lab frame (i.e., the external source frame of the AGN). 
}

\section{\label{sec:Geq}$\Gamma_{bulk}$ and equilibrium}

We assume a static bulk of relativistic leptons following an isotropic distribution in an anisotropic photon field. Owing to the Doppler effect, particles moving towards the main light source will scatter photons of higher energy and with a higher rate than those moving outwards. This will naturally lead to an anisotropic inverse Compton emission, most of it going back to the main photon source. This anisotropic emission will result in a transfer of momentum on the emitting plasma, the so-called Compton rocket effect first described  by \cite{1981ApJ...243L.147O}. A hot plasma could be driven to relativistic bulk motion through this mechanism. The force depends on the anisotropy of the soft photons seen in the comoving frame. When relativistic motion is taken into account, the photon field in the comoving frame will be affected by the bulk Doppler factor, resulting in a more isotropic photon distribution, until the Compton rocket force vanishes. The plasma then reaches an equilibrium velocity, or equivalently, an equilibrium bulk Lorentz factor $\Gamma_{eq}$.\\

In the Thomson regime, $\Gamma_{eq}$ depends only on the external radiation field through the Eddington parameters $(J,H,K)$ 
\begin{align}
J & = \frac{1}{4\pi}\int I_{\nu_s} (\Omega_s) \dd \Omega_s \dd \nu_s \notag\\
H & = \frac{1}{4\pi}\int I_{\nu_s} (\Omega_s)  \mu_s \dd \Omega_s \dd \nu_s \label{eq:edd_param}\\
K & = \frac{1}{4\pi}\int I_{\nu_s} (\Omega_s) \mu_s^2 \dd \Omega_s \dd \nu_s \notag
\end{align}

with $I_{\nu_s}$ the specific intensity of the emitting source, $\Omega_s$ the solid angle, and $\theta_s = \arccos \mu_s$ the angle under which the source is seen by the pair plasma (see figure \ref{fig:sketch}).\\

In the Thomson regime, the saturation of the Compton rocket effects happens when the second Eddington parameter, $H'$, vanishes  in the comoving frame (see \citealt{1997A&A...323..271M} for more details). With the factor $\displaystyle \zeta = \frac{J+K}{2H}$, one obtains the equilibrium equation
\begin{equation}
H' = H \ \Gamma_{eq}^2 \left( \beta_{eq}^2 - 2\zeta \beta_{eq} +1 \right)= 0
\label{eq:H'}
,\end{equation}
whose solution is
\begin{equation}
\beta_{eq} = \zeta - \sqrt{\zeta^2 -1}
\label{eq:Beta_eq_sol}
.\end{equation}

It is interesting to note again that, as long as the plasma is hot, this result does not depend on the jet model in the Thomson regime, but only on the external photon field (see equation \ref{eq:H'}). The Compton rocket effect will also take place in the Klein-Nishina (KN) regime, but in this case the computation of the equilibrium velocity is a bit more complex and depends on the  energy distribution of the pair (see \cite{1998MNRAS.300.1047R}). 
Moreover, one can expect that the resulting $\Gamma_{eq}$ will not vary much from that computed in the Thomson regime as long as we are not deep in the KN regime, as the recoil in the Thomson regime is much more efficient. Then $\Gamma_{eq}$ could be sensitive to KN correction for the most extreme objects ($>100$GeV; see Sect. 7).  In the following, we  suppose that conditions always meet the Thomson regime. 
The goal of this paper is to compute the resulting equilibrium bulk Lorentz factor $\Gamma_{eq}$, in a complex environnement, taking into account the angular and spectral distribution of various sources of photons in a realistic model of AGNs.\\

\section{\label{sec:modeling}Modeling the AGN}

\begin{figure}[ht]
\includegraphics[width=\hsize,trim= 3cm 2cm 3cm 0cm,clip=true]{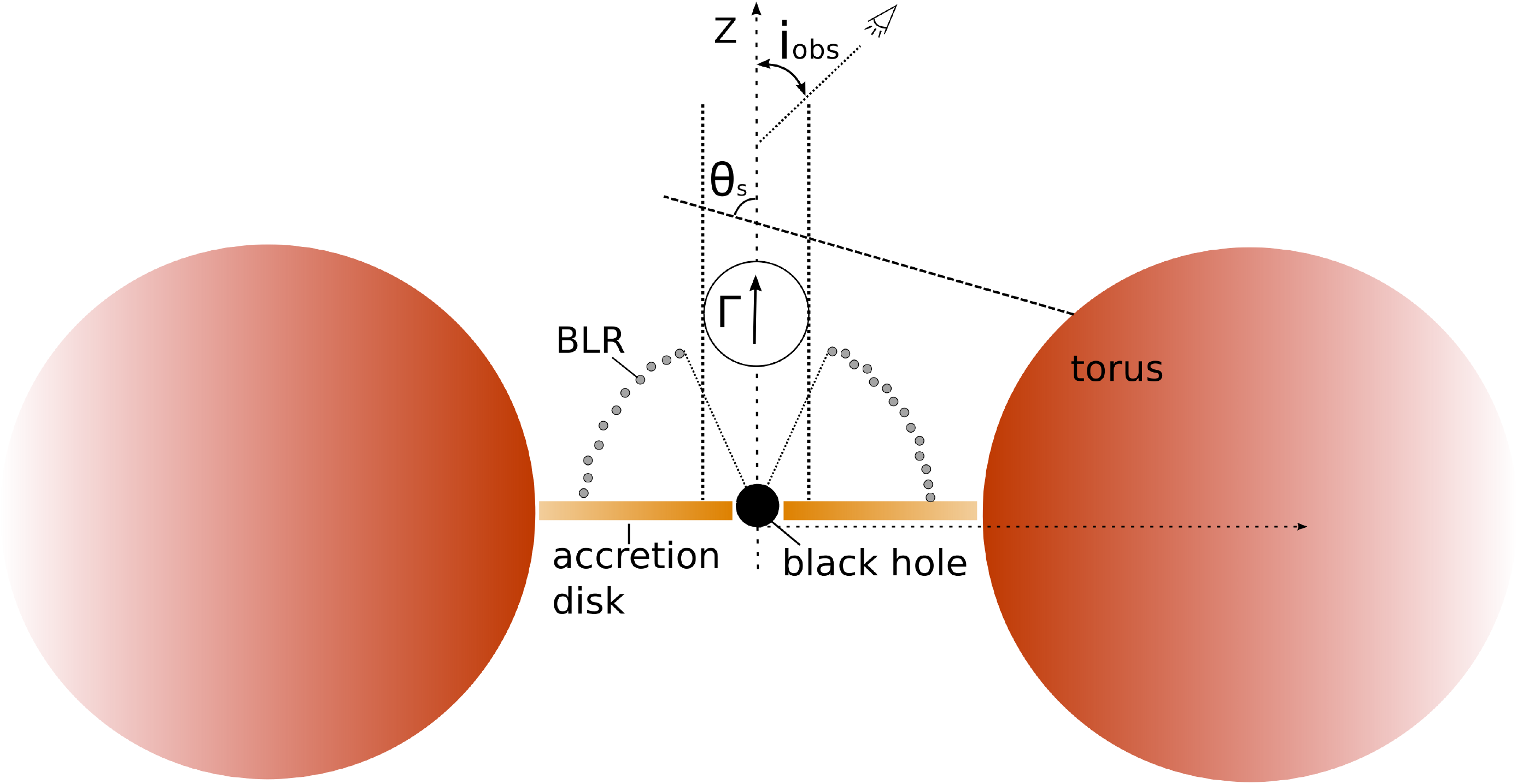}
\caption{\label{fig:sketch}The big picture:  edge-on view of the global model geometry (not to scale) with the accretion disk, the dusty torus, and the broad line region (BLR). $i_{obs}$ is the observer's viewing angle and $\theta_s$ is the angle between the incoming radiation from a source and the jet axis.}
\end{figure}

We will consider the effect of several possible external sources of photons, namely the accretion disk, the dusty torus, and the broad line region (see figure \ref{fig:sketch}). We note that because the synchrotron radiation is produced in the comoving frame with a zero net flux, it does not interfere with the bulk motion as long as the SSC is treated in a local approximation. This could  change, however,  if the particles scatter synchrotron photons produced in other parts of the jet, but this problem is much more complex since it involves the knowledge of the whole structure of the jet. We will not address this issue in this work.

\subsection{Discretization of the sources}

The anisotropy of the photon sources will be taken into account in the numerical scheme by slicing the different sources into a set of small independent parts modeled as graybodies in radiative equilibrium, i.e., with a Planck spectrum but with a possible smaller emissivity. This discretization is done in three dimensions. Even for an axisymmetric source, an azimuthal discretization is still required to  accurately compute the Compton external emission towards the observer's line of sight. Since the object is seen under a certain angle, the axisymmetry is always broken with respect to the line of sight.\\

Each slice of the discretization is then described by four numbers:
\begin{itemize}
\item $\mu_s = \cos \theta_s $ with $\theta_s$ the angle between the incoming light wave and the jet axis (see figure \ref{fig:sketch}),
\item $\dd \Omega$ the solid angle under which it is seen from the altitude $Z$,
\item its temperature $T$, 
\item its emissivity $\varepsilon$.
\end{itemize}

The slice emission characterized by these numbers is given by its specific intensity  $I_{\nu}$ $\left( erg \, s^{-1} \, cm^{-2} \, sr^{-1} \, Hz^{-1} \right)$, defined as the emitted energy $\dd E$ by normal surface $\dd \Sigma$, time $\dd t$, frequency band $\dd \nu$, and solid angle $\dd \Omega$: $dE = I_{\nu} \, \dd \Sigma \, \dd t \, \dd \Omega \, \dd \nu$ \citep{1979rpa..book.....R}. The specific intensity of a graybody is given by  Planck's law (equation \ref{eq:planck_law}):
\begin{equation}
\label{eq:planck_law}
 I_{\nu_s} =  \varepsilon \frac{2h\nu^3}{c^2}\frac{1}{\exp\left(\frac{h\nu}{k_B T}-1\right)}
.\end{equation}

A blackbody is a graybody with an emissivity $\varepsilon=1$.\\

Here below, we detail the computation of each photon source.
\subsection{Standard accretion disk}

The accretion disk is assumed to be an optically thick standard accretion disk as described by \cite{1973A&A....24..337S} extending from $R_{in}$ to $R_{out}$. Each point of the disk is a blackbody with a temperature given by the distance to the central black hole (BH), assumed to be non-rotating, following the relation

\begin{equation}
\label{eq:Tdisk}
T_{disk}(r) = \left[  \frac{3 G M \dot{M}} { 8 \pi \sigma} \frac{1}{r^3} \left(1 - \sqrt{\frac{3R_S}{r}} \right)   \right] ^{1/4}
\end{equation}
with $G$ the gravitational constant, $\sigma$ the Stefan-Boltzmann constant, $M$ the BH mass, $\dot{M}$ the accretion rate, $\displaystyle R_S = 2 \frac{GM}{c^2}$ the Schwarzschild radius, and $r$ the distance from the center of the BH.\\

\begin{figure}[h]
\centering
        \includegraphics[width=\hsize]{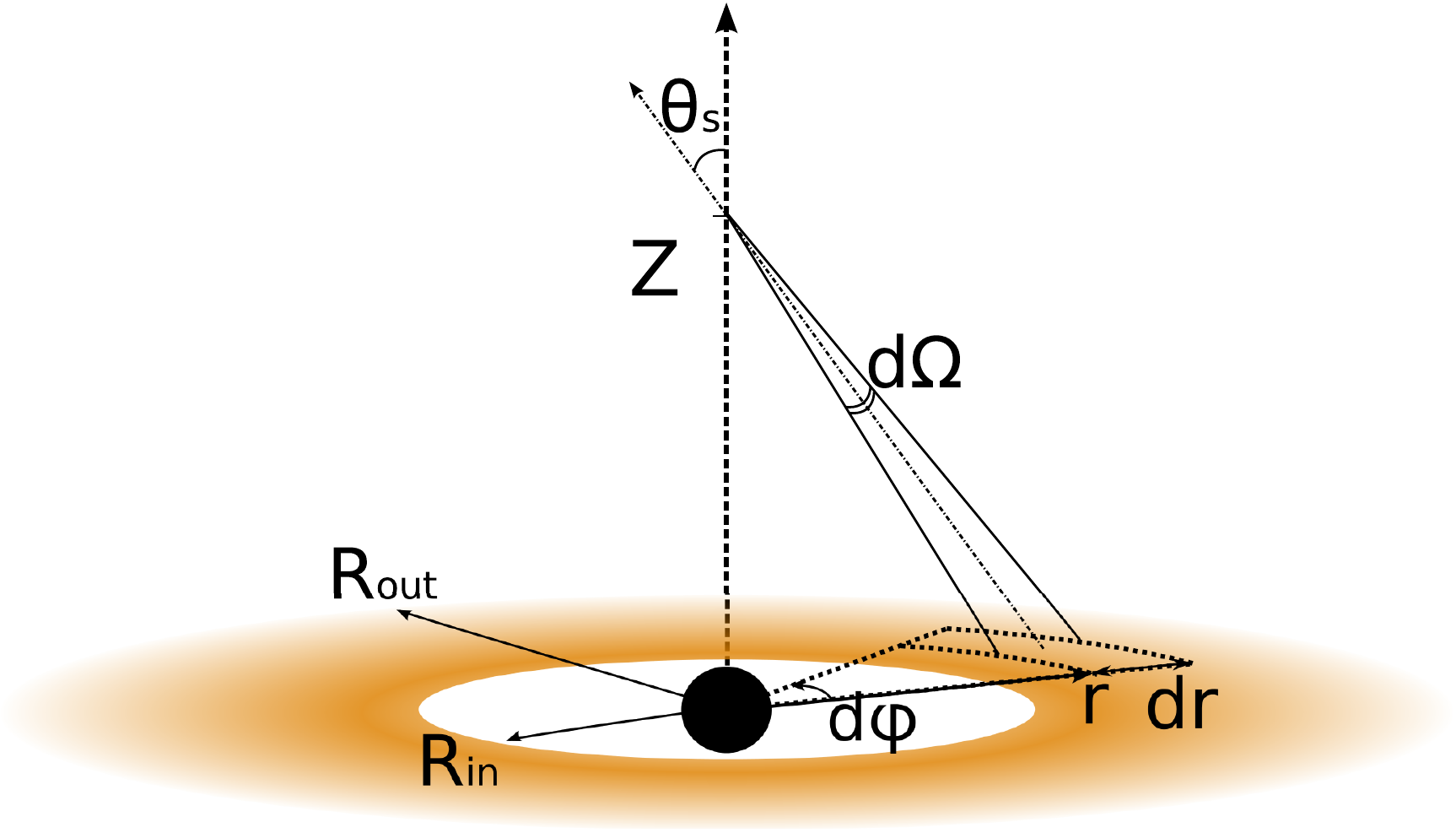}
        \caption{\label{fig:disk} Disk radial and azimuthal splitting. A slice at $(r,\varphi) \in \left( [R_{in},R_{out}],[0,2\pi]\right)$ is seen under a solid angle $\dd \Omega$ from the jet at an altitude $Z$.}
\end{figure}

To model this accretion disk, we sliced it in different parts, each being a blackbody with its own temperature $T_{disk}(r,\varphi)$. The discretization follows a logarithmic scale along $r$ and a linear scale along $\varphi$. Given an altitude $Z$ \LEt{ in the figure captions and in text below, this is given as upper case  Z   }in the jet, a slice of the disk is seen under a solid angle $\dd \Omega=Z \, \dd S /(r^2 + Z^2)^{3/2}$, where $\dd S = \dd \varphi (r dr +dr^2/2)$ is the surface of the slice. Photons coming from this slice make an angle $\theta_s = \arccos(Z / \sqrt{r^2+Z^2}) $ with the axis (see figure \ref{fig:disk}).

The luminosity of one face of the disk, for $R_{in} = 3R_S$ and $R_{out} \gg R_{in}$, is given by the relation
\begin{equation}
\label{eq:Ldisk}
L_{disk} = \int_{R_{in}}^{R_{out}} \sigma T^4_{disk}(r) \: 2 \pi r \: \dd r \approx \frac{\dot{M}c}{24}
.\end{equation}

\subsection{Dusty torus}

\begin{figure}[ht]
\includegraphics[width=\hsize]{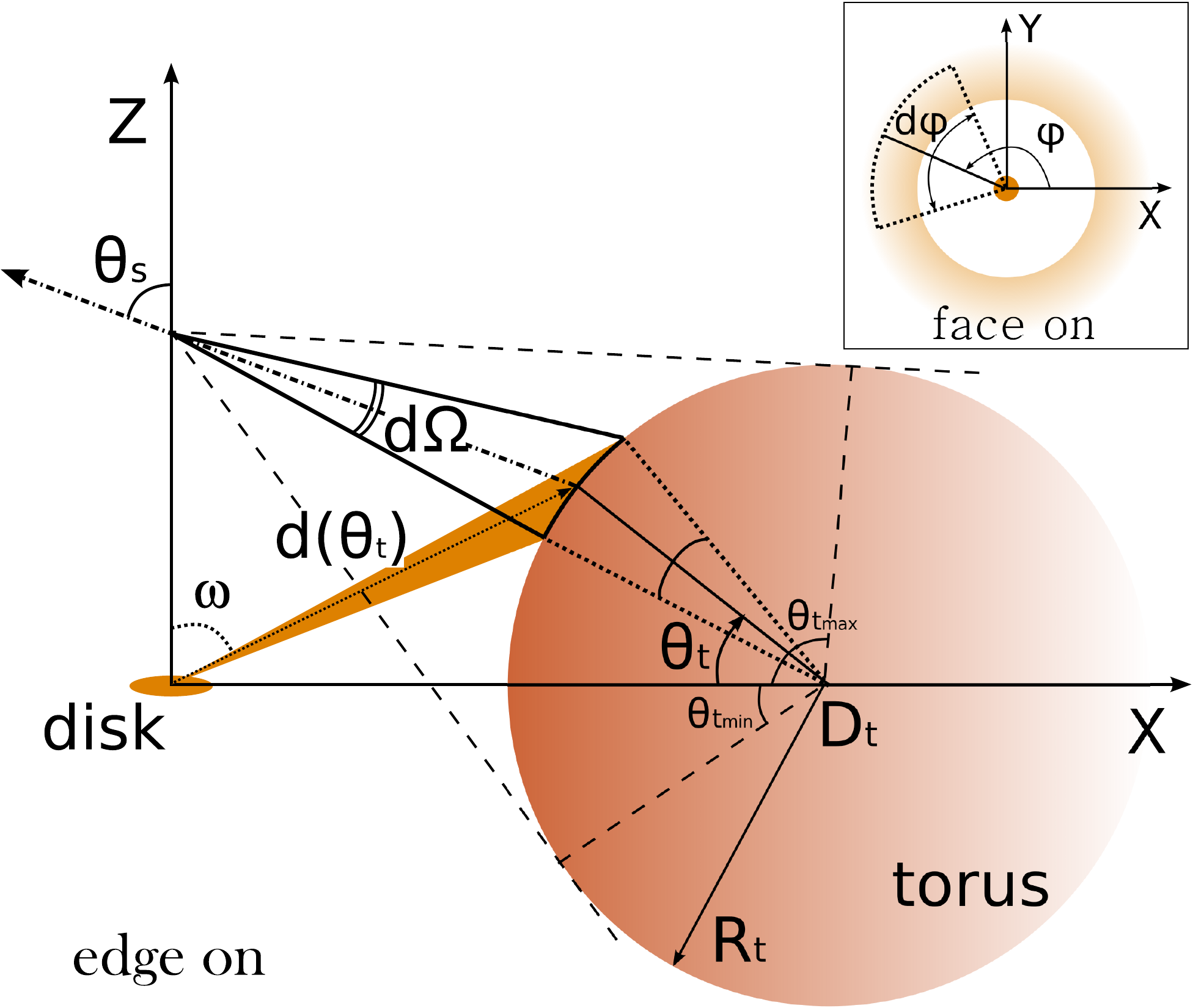}
\caption{\label{fig:torus}Dusty torus seen from an altitude Z in the jet under a solid angle $\dd \Omega$. The torus is sliced according to $\theta_t \in \left[\theta_{t_{min}},\theta_{t_{max}}\right]$ and $\varphi \in [0,2\pi]$. Each slice is illuminated by the disk and is in radiative equilibrium with a temperature $T(\theta_t,\varphi)$ and emits as a blackbody $(\varepsilon = 1)$.}
\end{figure}

The dusty torus is modeled by a torus shape structure whose major radius is $D_t$ and minor radius $R_t$ (see figure \ref{fig:torus}).

In the same way as the disk, the torus is sliced in different parts that will radiate as blackbodies (so $\varepsilon = 1$ for the torus). Slices follow a linear discretization and are located with their coordinates $\left( \theta_t,\varphi \right)$ at the surface of the torus.

Each slice of the torus is assumed to be in radiative equilibrium with the luminosity received from the accretion disk.
To simplify, we make the assumption that all the energy from the disk comes from its inner parts and that $R_{in} \ll \left(D_t-R_t \right)$ so that the source of energy is point-like when seen from the torus.
\corr{ With the parameter $ \displaystyle a=\frac{R_t}{D_t} \leq 1$, the torus temperature is given by

\begin{equation}
\label{eq:Ttorus}
T_{tor} (\theta_t) =
\left\{
      \begin{array}{l l}
         \displaystyle \left[ \frac{D_t(\cos\theta_t - a)}{2\pi\sigma d(\theta_t)^3}  L_{disk} \sin\omega \right]^{1/4}  & \text{\tiny for } \cos \theta_t \in \left[-a:a\right]\\
        0  & \text{otherwise}\\
      \end{array}
    \right.
\end{equation}
with $d(\theta_t)$ the distance between the slice center and the point-like source, 
\begin{equation}
\label{eq:d}
d(\theta_t) = D_t \left[ \left(1-a\cos\theta_t \right)^2+ a^2\sin^2\theta_t \right]^{1/2}
,\end{equation}
and $\omega$ the angle between the $Z$-axis and the emission direction from the disk:
\begin{equation}
\sin \omega  = \frac{a \sin\theta}{\left( 1+ a^2 -2a\cos\theta \right)^{1/2}}
.\end{equation}

}

From an altitude Z in the jet, the torus is seen under a certain solid angle which is delimited by $\theta_{t_{min}}$ and $\theta_{t_{max}}$ (see figure \ref{fig:torus}). These values can be determined from geometrical considerations:
\begin{align}
\theta_{t_{min}} & = \arctan\left({\frac{Z}{D_t}}\right) - \arccos\left({\frac{R_t}{\sqrt{Z^2 + D_t^2}}}\right),\\
\theta_{t_{max}} & = \arctan\left({\frac{Z}{D_t}}\right) + \arccos\left({\frac{R_t}{\sqrt{Z^2 + D_t^2}}}\right). \notag
\end{align}
However, in the case of a continuum between the accretion disk and the dusty torus ($D_t = R_t+r_{out}$), $\theta_{t_{min}}$ will be chosen equal to 0.

\subsection{Broad line region}

\begin{figure}[ht]
\includegraphics[width=\hsize]{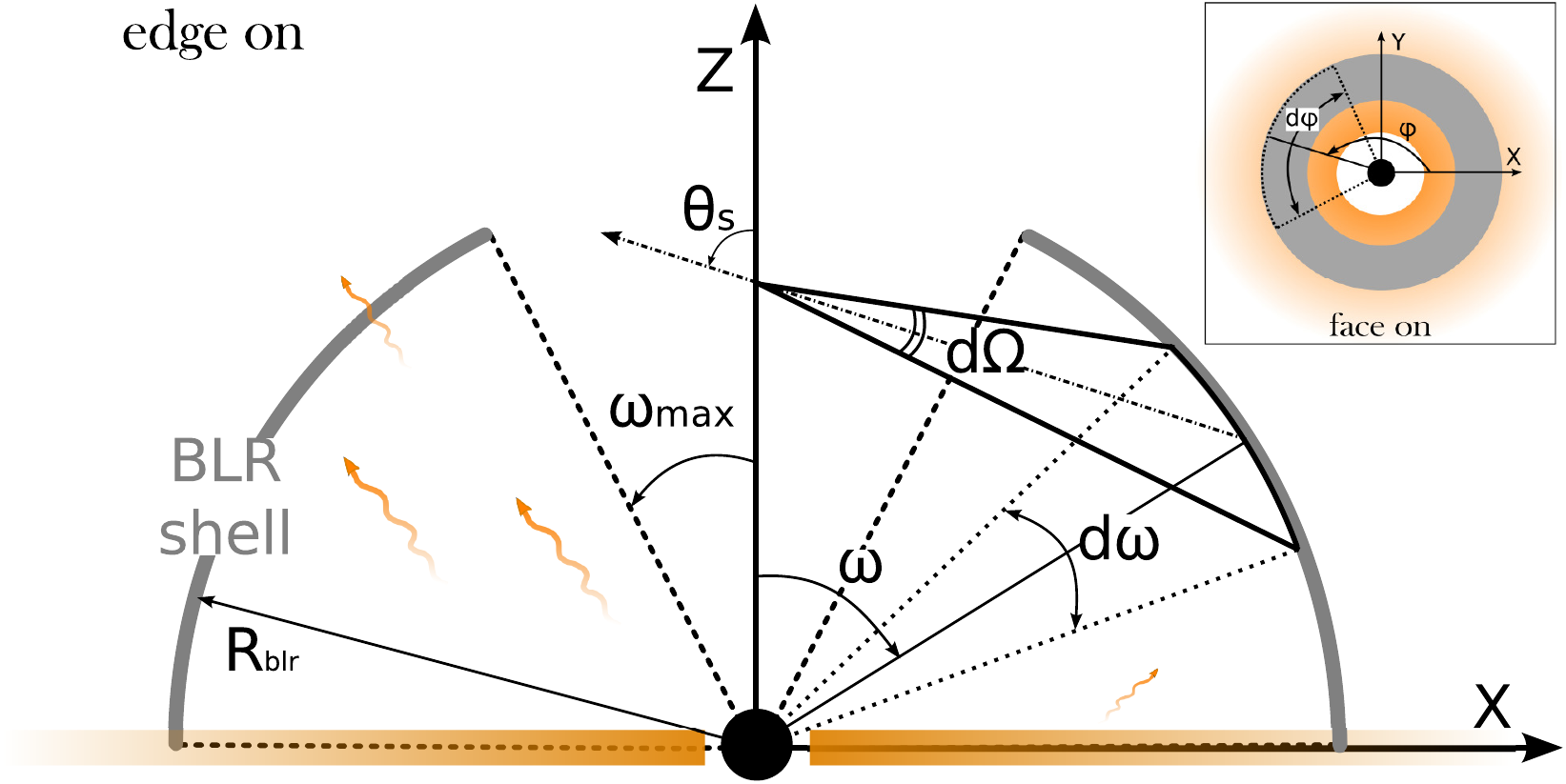}
\caption{\label{fig:blr}The BLR, an optically and geometrically thin shell of isotropic clouds seen from an altitude Z in the jet under a solid angle $\dd \Omega$. The BLR is sliced according to $\omega \in \left[\omega_{max},\pi/2\right]$ and $\varphi \in [0,2\pi]$. The BLR absorbs and re-emits part of the disk luminosity.}
\end{figure}

The broad line region is modeled as an optically and geometrically thin shell of isotropically emitting clouds situated at a distance $R_{blr}$ from the central black hole and extending up to an angle $\omega_{max}$ above the accretion disk plane (see figure \ref{fig:blr}).

\cite{Tavecchio:2008fu} showed that modeling the spectrum of the BLR with a blackbody spectrum at $T=10^5 K$ provides a good approximation of the resulting inverse Compton spectrum. We followed this idea using a temperature of $T_{blr} = 10^5 K$ and an overall luminosity being a fraction $\alpha_{blr}$ of the disk luminosity.
\corr{
To achieve this, the BLR is modeled as a graybody at $T_{blr}$ with an emissivity
\begin{equation}
\varepsilon_{blr} = \frac{\alpha_{blr} L_{disk}}{2\pi R_{blr}^2 \sigma T_{blr}^4} {\cos \omega_{max}}
.\end{equation}

Thus the total luminosity of the BLR is given by
\begin{equation}
\label{eq:Lblr}
L_{blr} = \int_{blr} \varepsilon_{blr}(\omega) \: \sigma T_{blr}^4 \dd S = \alpha_{blr}  L_{disk} 
.\end{equation}
}
Like the torus, the BLR is divided linearly into slices along $\omega$ and $\varphi$. 

\section{\label{sec:Geq_res}$\Gamma_{eq}$ in the jet}
\subsection{Parameter values}

We have presented  the   description of the source modeling, and we can now choose the values for the different parameters. They are listed in Table \ref{tab:sources_param}. If not specified otherwise, these parameters are set for the rest of the study.
Some characteristic values of the model are also derived in  table \ref{tab:sources_param}.

\begin{table}[ht]
\small
\begin{tabular}[width=\hsize]{l  l  l }
\hline
\hline\\[-6pt]
Parameter & Symbol & Value\\[2pt]
\hline\\[-6pt]
Black hole mass & M & $5 \times 10^8 M_{\odot}$\\
BH accretion rate & $\dot{M}$ & 1 $\dot{M}_{edd} $\\
Disk inner radius & $R_{in}$ & $3R_S$\\
Disk outer radius & $R_{out}$ & $5\times 10^4 R_S$\\
Disk emissivity & $\varepsilon_{disk}$ & 1\\
Number of disk slices &$N_{r_{disk}}\times N_{\varphi_{disk}}$& $18\times3$\\
Torus center & $D_t$ & $10^5 R_S$\\
Torus radius & $R_t$ & $5\times10^4 R_S$\\
Torus emissivity & $\varepsilon_t$ & 1\\
Number of torus slices & $N_{r_{tor}}\times N_{\varphi_{tor}}$ & $6 \times 3$\\
BLR radius &$R_{blr}$&$10^3 R_S$\\
BLR angular opening & $\cos\omega_{max}$ & 0.9\\
BLR temperature & $T_{blr}$ & $10^5K$\\
BLR absorption & $\alpha_{blr}$ & 0.1\\ 
Number of BLR slices & $N_{r_{blr}}\times N_{\varphi_{blr}}$ & $6\times3$\\[2pt]
\hline
\hline\\[-6pt]
Derived characteristic & Symbol & Value\\[2pt]
\hline\\[-6pt]
Schwarzschild radius & $R_S$ & $5.9 \times 10^{13} \, cm$ \\
Disk temperature & $T_{disk}$ &$[280 : 10^6]$ K\\
Disk total luminosity & $L_{disk}$ &  $1.0 \times10^{46} \, erg.s^{-1}$\\
Torus equilibrium&$T_{tor}$& $[580 : 1345]$ K\\
 temperature &&\\
Torus total luminosity& $L_{tor}$& $1.4\times10^{45} \, erg.s^{-1}$\\
BLR total luminosity & $L_{blr}$ & $1.0\times10^{45} \, erg.s^{-1}$ \\[2pt]
\hline
\end{tabular}
\caption{\label{tab:sources_param}Parameters of the external sources. The values indicated in the right column are those used in this paper.}
\end{table}

\begin{figure}[ht]
\centering
        \includegraphics[width=\hsize]{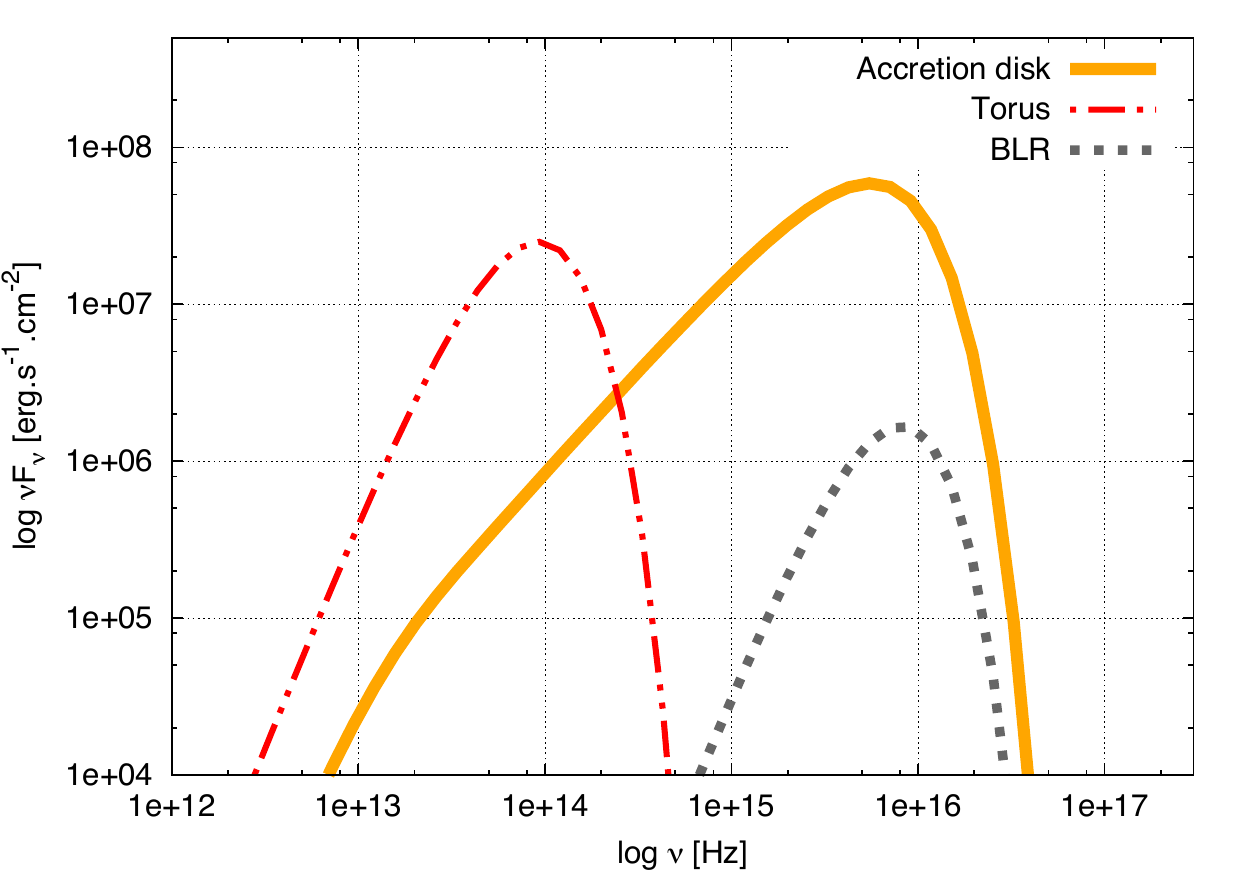}
        \caption{\label{fig:sources_spec} Spectra of the three external soft photons sources seen at an altitude $Z=10^5 R_S$ on the jet axis in the sources frame.}
\end{figure}

With these  parameters the external source spectra can be derived. An illustration of these spectra seen at an altitude $Z=10^5 R_s$ in the jet is provided in figure \ref{fig:sources_spec}.

\subsection{Resulting $\Gamma_{eq}$ \label{sec:gam_eq}}

\begin{figure}[ht]
\centering
        \includegraphics[width=\hsize]{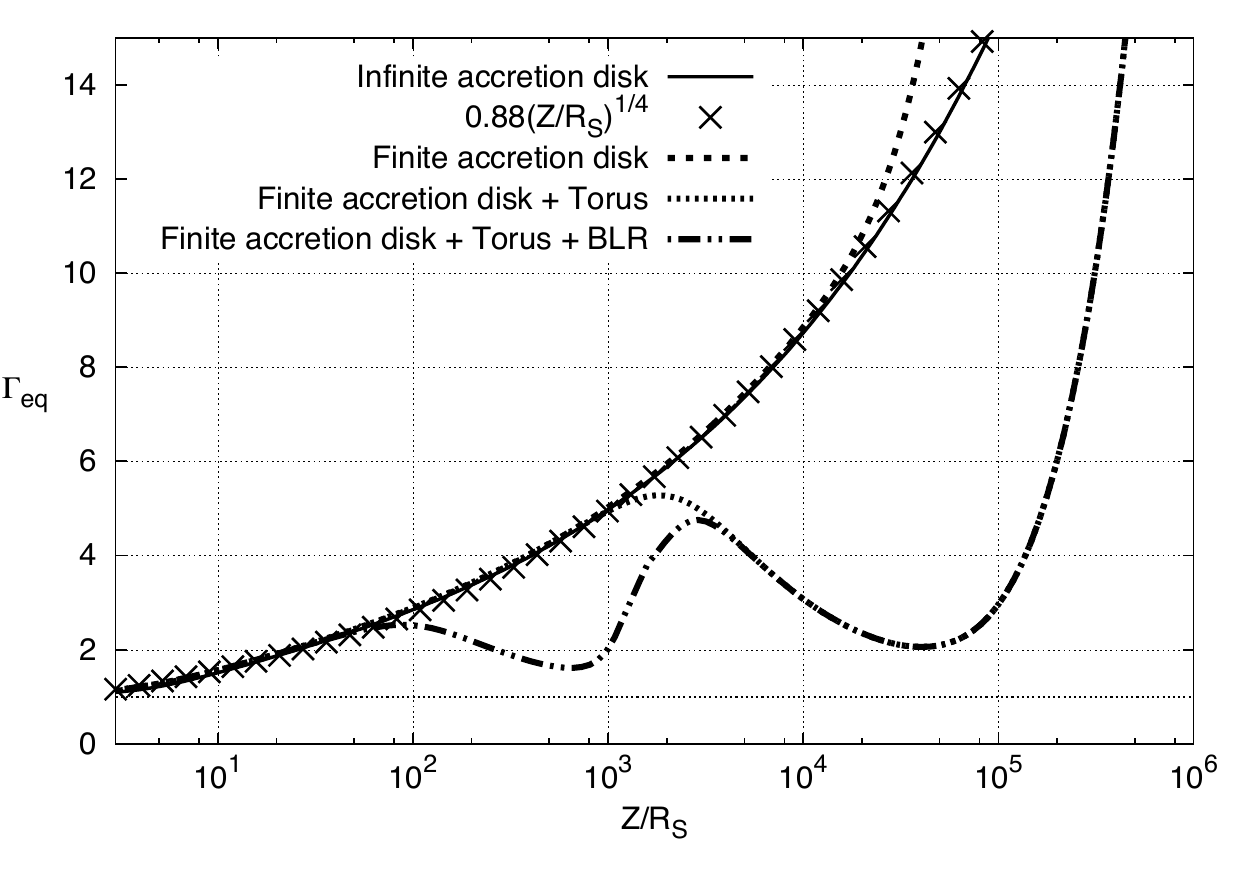}
        \caption{\label{fig:gam_eq} Equilibrium bulk Lorentz factor resulting of the external Compton emission for different external photon sources. The geometry is described in figure \ref{fig:sketch} with the following parameters: finite and infinite accretion disk have an inner radius $R_{in}=3 R_S$; the finite disk has an outer radius $R_{out}=5 \times10^4 R_S$; $D_{torus} =10^5 R_S$, $R_{torus} = 5 \times10^4 R_S$, $R_{BLR} = 10^3 R_S$, and $\cos\omega_{max}=0.9$.}
\end{figure}

We can now compute the resulting equilibrium bulk Lorentz factor $\Gamma_{eq}$ all along the jet. This has been done for different set-ups of external sources (infinite accretion disk or finite accretion disk alone, finite disk + torus or finite disk + torus + BLR) and the results are given in figure \ref{fig:gam_eq}.
In this plot, one can distinguish the effect of each external source.

We start with the case of an infinite accretion disk $(R_{out} = \infty)$. For an emission zone in the jet, the inner parts of the accretion disk are always seen from below. As explained earlier, for a static source, this would lead to a Compton emission mainly directed toward the disk. This causes a transfer of momentum and a thrust forward on the plasma. However, as soon as the plasma accelerates, the photons coming from the outer part (for $\mu_s < \beta^{-1}$) of the disk seem to travel backward in the comoving frame and produce a drag on the plasma. At every altitude, the equilibrium velocity is reached when the two effects balance. Analytical computation from \cite{1995MNRAS.277..681M} showed that in the case of an infinite accretion disk, one should have $\Gamma_{eq} = 1.16 \left( \frac{Z}{R_i} \right)^{1/4}$. With $R_i=3R_S$, this gives $\Gamma_{eq} = 0.88 \left( \frac{Z}{R_S} \right)^{1/4}$, which is in agreement with our numerical results (see bold solid line compared to crosses in figure \ref{fig:gam_eq}).

If we now consider  a finite accretion disk $(R_{out} = 5 \times 10^4 R_S)$, we note the same behavior: at low altitudes the disk seems to be infinite seen from the axis. Once  an altitude $Z \gtrsim R_{out}$ is reached, the drag effect from the outer parts of the disk ceases and then the entire disk will imply a thrust on the bulk. As long as the acceleration is effective, $\Gamma_{eq}$ will follow a law in $Z/R_{out}$ (\cite{1998MNRAS.300.1047R}).\\
\corr{
We can have the same reasoning concerning the effect of the dusty torus. As seen from the jet axis, the radiation from the torus comes at greater angles than the one from the accretion disk. \LEt{ yes? }Therefore, when the plasma accelerates, the torus radiation seems to come forward, which will tend to drag the flow. Nevertheless, in the lowest altitude the accretion disk radiation dominates and the resulting $\Gamma_{eq}$ is unchanged from the previous case. It is only from a certain altitude ($Z \approx 10^3 R_S$ in our study) that the effect of the torus radiation starts to dominate and that the flow will actually slow down. 
Of course, the equilibrium velocity will never reach zero, as the radiation in the lab frame is never isotropic and always has a preferred direction upward. At one point ($Z \approx R_t$ in our study), most of the radiation from the torus moves \LEt{ ok? } forward in the comoving frame. This leads to a thrust on the flow and $\Gamma_{eq}$ increases again with the same accelerating slope as in the finite accretion disk case.

The BLR photon field shows the same kind of effects with a deceleration regime inside the BLR (from $10^2 R_S$ to $10^3 R_S$ in our study), followed by an accelerating regime once the bulk leaves the BLR. At some point, the torus photon field becomes predominant and controls $\Gamma_{eq}$ as explained previously.\\

We note that the computation is done for a hot electron-positron plasma in the Thomson regime. For very hot plasmas, KN corrections will affect the rate of momentum transfer and the bulk Lorentz factor will differ in a way that is difficult to predict. Indeed, the Compton rocket is less efficient in the KN regime, but photons coming from larger incident angles are more likely to be in the KN regime and are precisely the ones dragging the flow.

Moreover, far from the external photon sources, the relaxation time to the equilibrium will become larger than the dynamical time $z/c$. At this point, the acceleration will stop, leading to an asymptotic value of the bulk Lorentz factor. The blob will then follow a ballistic motion. However, The point where this decoupling occurs depends on the absolute luminosity of the disk and the average energy of the plasma. A study of this phenomenon is presented in section \ref{sec:Ginf}.
}
\section{\label{sec:param_evol_gam}Influence of parameters on the equilibrium bulk Lorentz factor}

In this section we  study the influence of the model parameters on $\Gamma_{eq}$ with the model composed of the accretion disk, the dusty torus at thermal equilibrium, and the BLR. If not  otherwise stated, the parameters keep the values given in the previous section (table \ref{tab:sources_param}).

\subsection{Influence of the BLR on $\Gamma_{eq}$}

\subsubsection{Influence of the BLR opening angle $\omega_{max}$}

\begin{figure}[ht!]
\centering
        \includegraphics[width=\hsize]{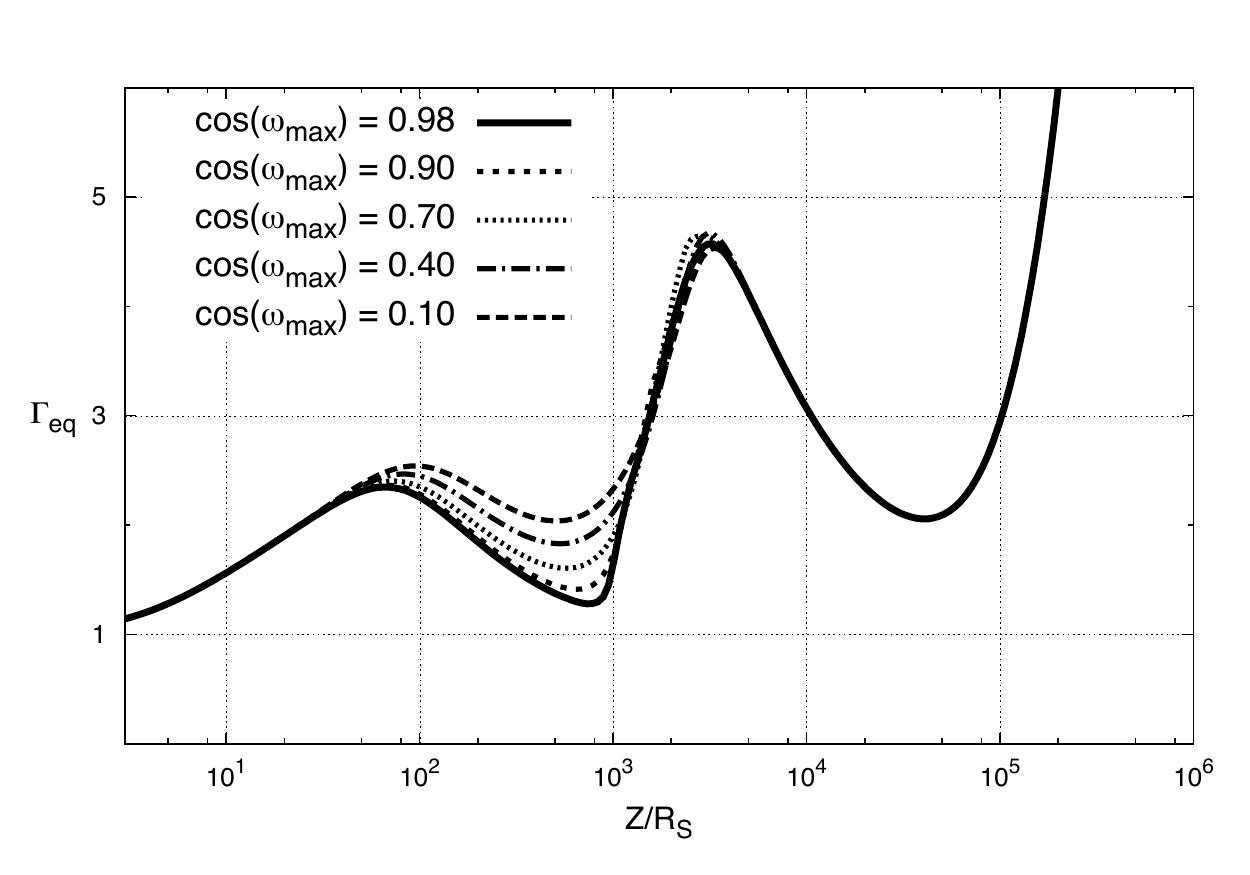}
        \caption{\label{fig:GAMcosomegamax} Equilibrium bulk Lorentz factor for several BLR opening angles $\cos \omega_{max}$. Values of the other geometrical parameters can be found in table \ref{tab:sources_param}.}
\end{figure}

The influence of the BLR opening angle on $\Gamma_{eq}$ is shown in figure \ref{fig:GAMcosomegamax}. The smaller $\omega_{max}$, the bigger the BLR, and the stronger the effect.\\

\corr{
When $\omega_{max}$ increases, parts of the BLR at small $\omega$ are suppressed.
For the plasma inside the BLR ($Z<R_{blr}$), radiation from these parts  moved backward in the comoving frame. The suppression of this radiation means less dragging effect and thus a higher $\Gamma_{eq}$ for the flow inside the BLR. Thus, the differences between the different opening angles are important at $Z<R_{blr}$

However, for the plasma outside the BLR ($Z>R_{blr}$), radiation from these parts moved forward in the comoving frame. The suppression of this radiation means less thrust on the plasma,  but the radiation from the parts of the BLR at greater $\omega$, which seems to move forward in the comoving frame, is still present and so is the dragging effect. This is why the differences between the different cases at $Z>R_{blr}$ are not so important. Much of the thrust is given \LEt{ ok? I had to rearrange a little in the previous paragraphs. Have I interpreted correctly? } by the disk itself, even at altitudes close to the outer border of the BLR.
}

\subsubsection{Influence of the BLR absorption $\alpha_{blr}$}

\begin{figure}[ht!]
\centering
        \includegraphics[width=\hsize]{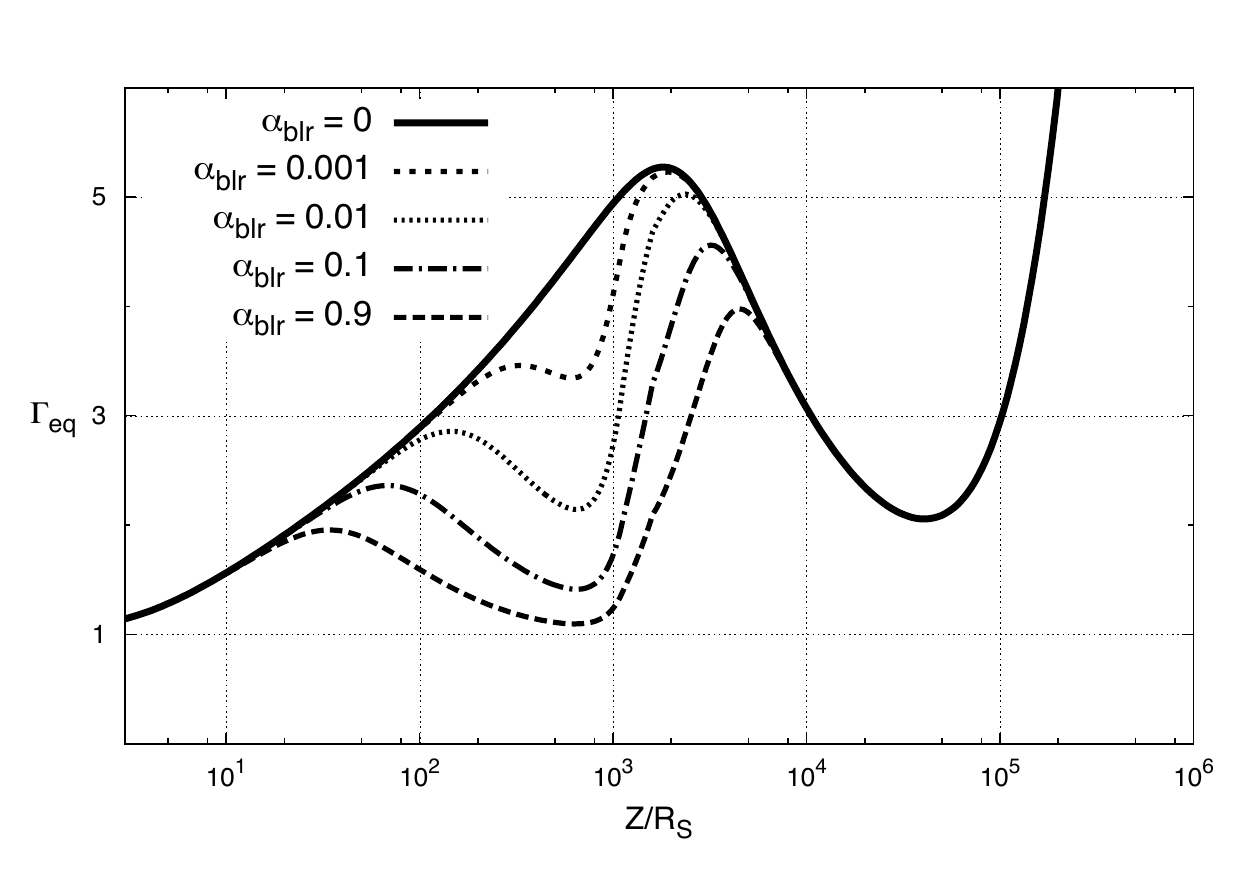}
        \caption{\label{fig:GAMabsblr} Equilibrium bulk Lorentz factor for several BLR absorption $\alpha$. Values of the other geometrical parameters can be found in table \ref{tab:sources_param}.}
\end{figure}

The effect of the BLR absorption $\alpha_{blr}$ is shown in figure \ref{fig:GAMabsblr}. The effect is  similar to the opening angle effect. The stronger the absorption, the stronger the emissivity and thus the stronger the luminosity from the BLR, and consequently, the stronger the drag effect.

However, unlike in the opening angle $\omega_{max}$ case, differences in the acceleration regime outside the BLR are noticeable. Indeed, with a reduction of the absorption, the drag and the thrust  change, which was not the case previously.

\subsubsection{Influence of the BLR radius}

\begin{figure}[ht]
\centering
        \includegraphics[width=\hsize]{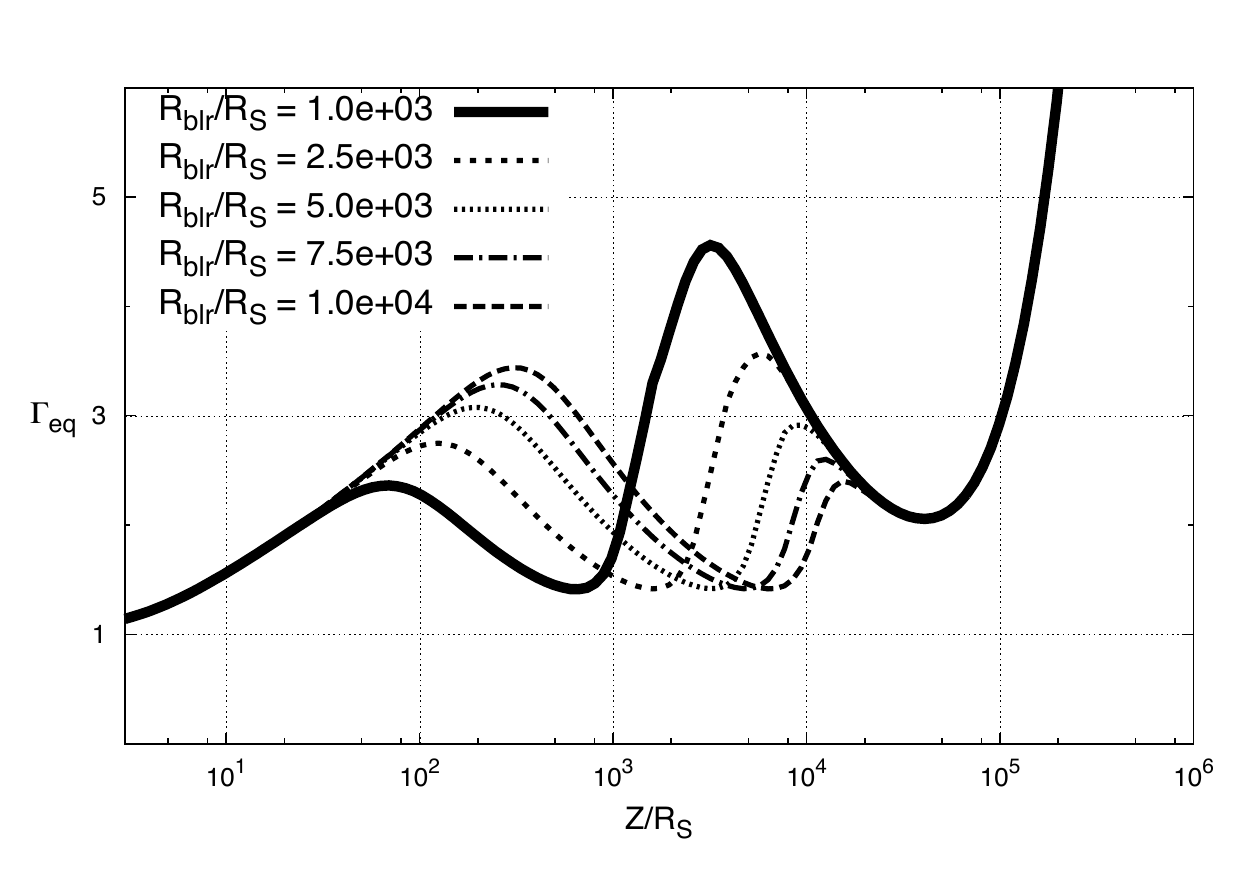}
        \caption{\label{fig:GAMRblr} Equilibrium bulk Lorentz factor for several BLR radius $R_{blr}$.  Values of the other geometrical parameters can be found in table \ref{tab:sources_param}.}
\end{figure}

The effect of the BLR radius is shown in figure \ref{fig:GAMRblr}. The BLR radius is now going from $R_{blr}=10^3 R_S$ to $R_{blr} = 10^4 R_S$. Other parameters are fixed to the values given in table \ref{tab:sources_param}.

The effect of the source radius sizes is a bit different as it  increases the amplitude of the drag or thrust, but also shifts the different regimes in altitude \LEt{ ok? "not only...but also" is a heavy construction, to be avoided when possible    }. We  note that when the BLR radius increases, its acceleration zone moves to higher $Z$. At one point (when $R_{blr}$ tends to $D_t-R_t$), the radiation from the torus dominates and  controls $\Gamma_{eq}$.

\subsection{Influence of the torus on $\Gamma_{eq}$}

\begin{figure}[ht]
\centering
        \includegraphics[width=\hsize]{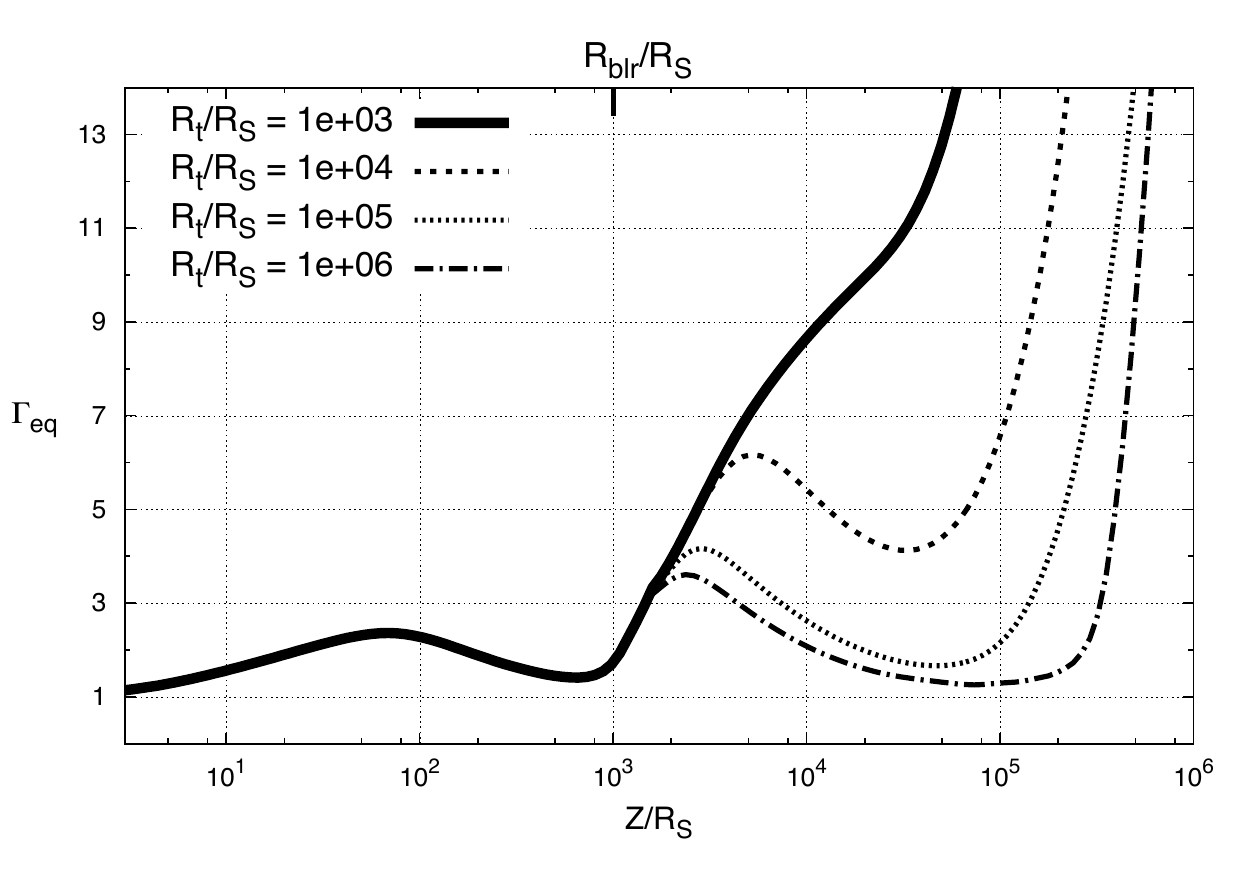}
        \caption{\label{fig:GAMRtorus} Equilibrium bulk Lorentz factor for several torus sizes. $R_t$ changes as a free parameter as $D_t = r_{out} + R_{t}$.  The values of the other parameters can be found in table \ref{tab:sources_param}.}
\end{figure}

We study here the influence of varying $R_t$, but always assume that $D_{t}=r_{out}+R_t$. The results are shown in figure \ref{fig:GAMRtorus}.

The torus acts farther \LEt{ farther,  i.e., distance? or to a greater degree? Perhaps: the torus has a greater/stronger effect on the jet than the BLR does.  }in the jet than does the BLR. Its radius effect is similar to the BLR's: the bigger the radius, the stronger the drag, and the  lower  in the jet it occurs. With a smaller torus, higher $\Gamma_{eq}$ are reached closer to the black hole.
Therefore, the emission from the lowest parts of the jet (at subparsec scale) will be strongly influenced by the torus size through the induced Doppler boosting.

\subsection{\label{sec:iobs}Observation angle and Doppler factor}

The relativistic bulk Doppler factor is defined as
\begin{equation}
\delta_b = \frac{1}{\Gamma_b \left( 1 - \beta_b \, \mu_{obs}\right)}
\end{equation}
with $\mu_{obs} = \cos i_{obs}$ (see figure \ref{fig:sketch} for a definition of $i_{obs}$).
As $\displaystyle \frac{I_\nu}{\nu^3}$ is a relativistic invariant \citep{1979rpa..book.....R}, the specific intensity in the lab frame is given by $I_\nu = I'_\nu \, \delta_{b}^3$. It can be  shown that most of the emission is emitted within a characteristic emission cone of aperture angle $\approx 1/\Gamma$.
This led to the idea that the same object seen from a different angle will show a different broadband spectrum and led to the AGN unification scheme (\citealt{Blandford:1978un}, \citealt{Orr:1982vv} and \citealt{1989ApJ...336..606B}).

\begin{figure}[ht]
\centering
        \includegraphics[width=\hsize]{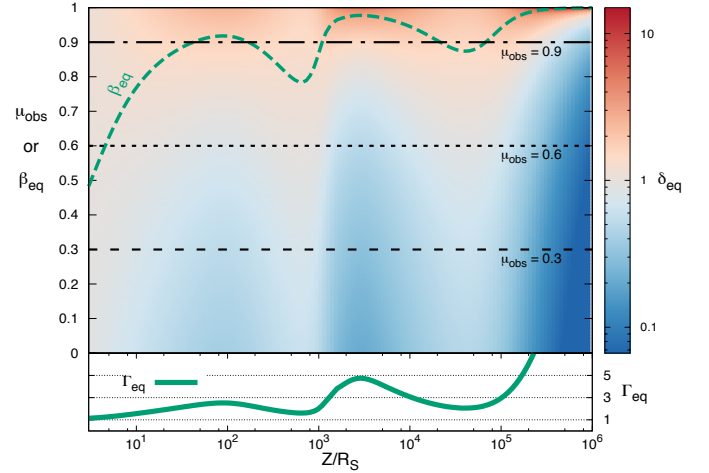}
        \caption{\label{fig:deltaobs_colors} Upper panel: Equilibrium bulk Doppler factor $\delta_{eq}$ in the jet plane altitude $Z/R_s$ vs observational angle $i_{obs}$ ($\mu_{obs} = \cos i_{obs}$).  The color scale is shown on the right. The green dashed line represents the corresponding $\beta_{eq}$. Bottom panel: $\Gamma_{eq}$ as a  function of the jet altitude. The geometry is described in section \ref{sec:modeling} and the source parameters are listed in table \ref{tab:sources_param}.}
\end{figure}

\begin{figure}[ht]
\centering
        \includegraphics[width=\hsize]{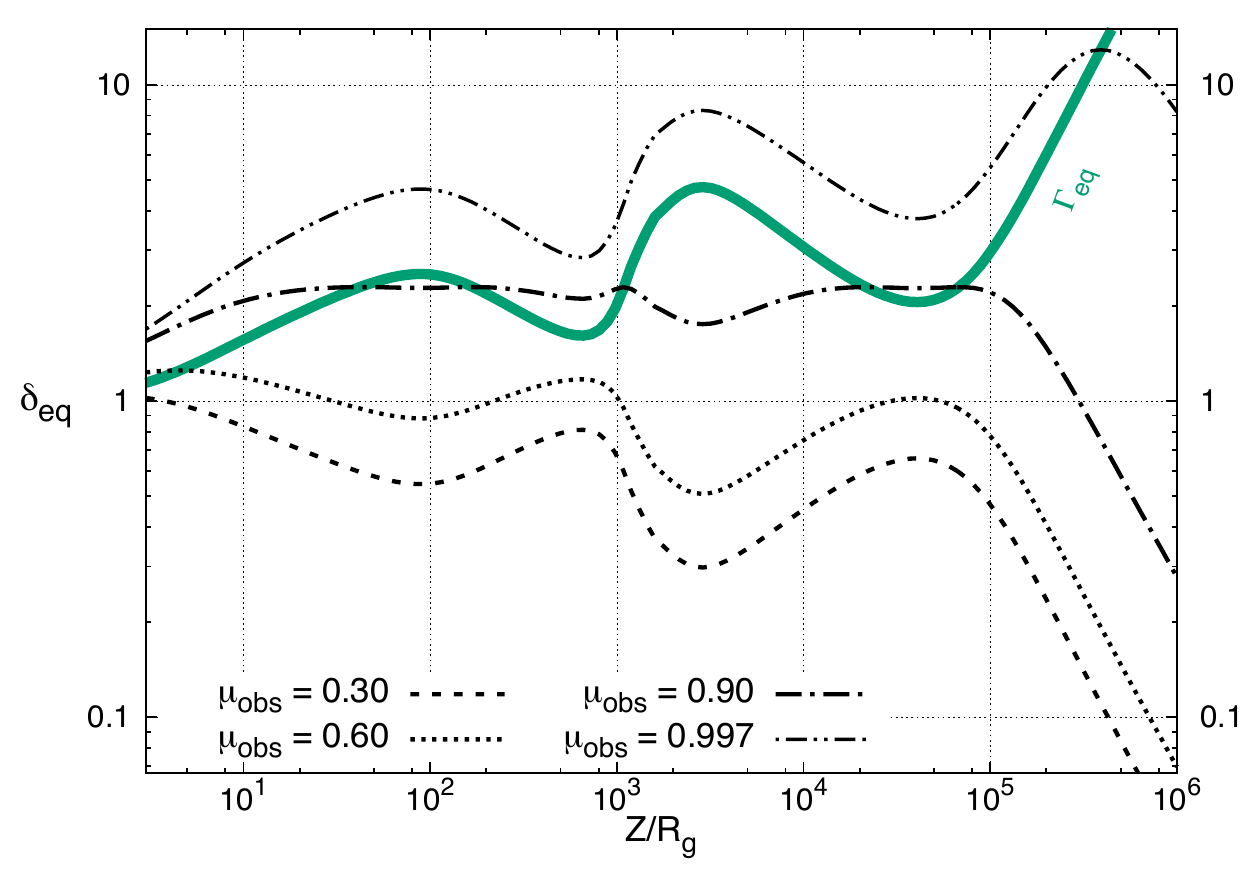}
        \caption{\label{fig:deltaobs} Examples of equilibrium bulk Doppler factor as a function of the altitude for different observational angles $i_{obs}$ $\left( \mu_{obs} = \cos i_{obs}\right)$. The green solid line represents the corresponding $\Gamma_{eq}$. The geometry is described in section \ref{sec:modeling} and the source parameters are listed in table \ref{tab:sources_param}.}
\end{figure}

For a given function $\Gamma_{eq}(Z)$, it is possible to  compute the function of the equilibrium bulk Doppler factor $\delta_{eq}(Z, \mu_{obs}),$ which depends on the altitude and on the observer viewing angle.
Figure \ref{fig:deltaobs_colors} shows the function $\delta_{eq}(Z,\mu_{obs})$ in false colors, corresponding to the $\Gamma_{eq}$ (see also   figure \ref{fig:deltaobs_colors}) computed in section \ref{sec:Geq_res} with the source parameters given in table \ref{tab:sources_param}. Figure \ref{fig:deltaobs} shows the same $\delta_{eq}$ as a  function of the altitude, but for four chosen observation angles.

An observer situated at a constant $\mu_{obs}$ sees the emission along the jet modulated by $\delta_{eq}(Z)$. It can be seen in  figures \ref{fig:deltaobs_colors} and \ref{fig:deltaobs} that $\delta_{eq}(Z)$ shows several extrema at a constant $\mu_{obs}$. This means that certain zones of the jet are preferentially seen depending on the jet viewing angle $i_{obs}$.

A few remarks can be made for peculiar values of $\delta_b$.
\vspace{-1ex}
\paragraph{a.} 
$\delta_b = 1$ is an important value for the observer because it marks the limit between an increase and a decrease in the observed flux compared to the flux in the comoving frame. As $\Gamma_{eq}(Z)$ varies in the observer frame, so does $\delta_{eq}$, and the same observer can be in the emission cone of certain parts of the jet and out of the emission cone of other parts of the jet. With the bulk Lorentz factor computed in section \ref{sec:Geq_res}, a value of $\delta_{eq} = 1$ is possible for $0.52 < \mu_{obs} < 0.96$.

\paragraph{b.}
Extrema of $\delta_{eq}$ are found at Z verifying
\begin{equation*}
\frac{\partial \delta_{eq}}{\partial Z} = 0 \Leftrightarrow  \left. \frac{\partial \delta_{b}}{\partial \Gamma} \right |_{\Gamma_{eq}} \left( \frac{\partial \Gamma_{eq}}{\partial Z} \right) = 0.
\end{equation*} 
We are then left with two possibilities:
\begin{itemize}
\item[$\bullet$] $\displaystyle \frac{\partial \Gamma_{eq}}{\partial Z}= 0$\\
The solutions to this equation correspond \LEt{ The  altitudes found using this equation correspond ? }  to the $\Gamma_{eq}$ extrema. In our case, they are approximately at $Z = 134 \, R_S$, $Z = 656 \, R_S$, $Z = 1.73 \times 10^3 \, R_S$, and $Z=3.61 \times 10^4 \, R_S$.\\

\item[$\bullet$] $\displaystyle \left. \frac{\partial \delta_{b}}{\partial \Gamma} \right |_{\Gamma_{eq}} =0$
\end{itemize}
It can be  shown that this always happens for $ \displaystyle \beta_{b} = \mu_{obs} $, or equivalently for $\displaystyle \delta_{b}=\Gamma_{b}$. These $\delta_{b}(Z)$ extrema can be assimilated to a way in or out of the emission cone by the observer at $\mu_{obs}$. In our particular case, figure \ref{fig:deltaobs_colors} shows that this is possible for $\displaystyle \text{min}(\beta_{eq}) = 0.45 \lesssim \mu_{obs} \lesssim 1 = \text{max}(\beta_{eq})$. Moreover, it  can be confirmed for \LEt{ ok? } the cases $\mu_{obs} = 0.6 - 0.9 - 0.997$ that the altitudes where $\beta_{eq}=\mu_{obs}$ in figure \ref{fig:deltaobs_colors} correspond to the altitudes where $\delta_{eq}$ reaches an extremum and where $\delta_{eq} = \Gamma_{eq}$ in figure \ref{fig:deltaobs}.

However,  $\Gamma_{eq}$ extrema can correspond to $\delta_{eq}$ minima or maxima \LEt{ plural? minima or maxima? }depending on $\mu_{obs}$.
In the case $\beta_{eq} < \mu_{obs}$ (blazar-type objects), $\delta_{eq}$ maxima are correlated with $\Gamma_{eq}$ maxima.
In consequence, for these objects an observer will  preferentially see jet zones where $\Gamma_{eq}$ is at a maximum since the jet emission will be more boosted. \LEt{ ok? }On the contrary, for $\beta_{eq} > \mu_{obs}$ (radio-galaxy-type objects), $\delta_{eq}$ maxima are correlated with $\Gamma_{eq}$ minima. This means that an observer will not see the zones of the jet that have the highest speed, but -- on the contrary-- the jet emission will be dominated by the slowest zones.

It is also interesting to note that there is a class of objects that will present very low modulation of the jet emission along $Z$.
These objects are characterized by $\displaystyle \mu_{obs} \approx \beta_{eq}$ and thus $\displaystyle \frac{\partial\delta_{eq}}{\partial Z} \approx 0$ almost everywhere in the jet. An example of this is shown by the case $\mu_{obs} = 0.9$ in figure \ref{fig:deltaobs}, where $\delta_{eq}$ is almost constant from $Z=10R_S$ to $Z=10^5R_S$ .  Of course other sources of variations can still produce \LEt{ still imply?  can still produce? }an important variability for these sources.\\

Similarly, different processes could dominate at different altitudes, only due to $\Gamma_{eq}$ evolution. In particular, external and synchrotron self-Compton emissions do not have the same beaming pattern for a given Doppler factor as shown by \cite{1995ApJ...446L..63D}. This author showed that synchrotron self-Compton follows a general beaming pattern $\propto \delta_b^{3+\alpha}$ (with $\alpha$ the energy spectral index of the radiation) whereas external Compton follows a beaming pattern $\propto \delta_b^{4+2\alpha}$  because the Comptonized photon field is isotropic in the plasma rest frame in the SSC case whereas it depends on $\delta_b$ in the external Compton case.

Moreover, the computation in \cite{1995ApJ...446L..63D} assumed an external isotropic radiation and a pre-assumed bulk Lorentz factor. Both of these assumptions are no longer valid  in our framework.
This could have some consequences on the beaming statistics of these objects, but an exhaustive study of these effects could not be done without a complete modeling of the jet, which is not the purpose of this paper.

\section{\label{sec:Ginf}Dependence of $\Gamma_b$ on the energetics}
\corr{
As stated previously, in the two-flow paradigm the Compton rocket process finds its energy in the turbulence from the outer MHD jet through the relativistic particle emission. It is therefore understandable that the energetics of the particles will limit the influence of the Compton rocket effect on the actual value of $\Gamma$. To compute the actual value of $\Gamma$, one can solve the following differential equation (\cite{1998MNRAS.300.1047R}),

\begin{equation}
\frac{\partial \Gamma_b(Z,\gamma_e)}{\partial Z} = \frac{F^{\prime z}} {\rho'} \frac{1}{\left(1+\frac{1}{3\Gamma_b^2}\right)}
\end{equation}

with $\displaystyle F^{\prime z} = \frac{\sigma_T}{c} 4 \pi H' \int \left( 1+\frac{2}{3}\gamma^{\prime 2}_e \beta^{\prime 2}_e \right) n'_e(\gamma')\dd \gamma'$

and $\displaystyle \rho' = \int \gamma' m_e c^2 n'_e(\gamma') \dd \gamma'$.

Here, it can be seen that the complete calculation of $\Gamma_b(Z,\gamma_e)$ depends on the particle energy distribution.
For the sake of simplicity, here we choose a Dirac distribution:
\begin{equation}
n_e(\gamma) = N_e \, \delta\left(\gamma - \gamma_e \right).
\end{equation}

In this case,  the following results are obtained:
\begin{equation}
F^{\prime z} = \frac{\sigma_T}{c} \frac{8 \pi}{3} N_e \gamma_e^2 H' 
\end{equation}
and
\begin{equation}
\rho' = N_{e} \gamma_e m_{e} c^{2}
.\end{equation}

For bulk Lorentz factors close to the equilibrium value,  $H'$ can be evaluated with a linear expansion:
\begin{equation}
H'(\Gamma_{b}) \approx H'(\Gamma_{eq}) + \frac{\dd H'}{\dd \Gamma_{b}} \left( \Gamma_{b} - \Gamma_{eq}\right) 
.\end{equation}

By definition $\displaystyle H'(\Gamma_{eq}) = 0,$ and it can be shown that
\begin{equation}
H' \approx -\frac{H}{\beta_{eq}^{3}\Gamma_{eq}^{3}} \left( \Gamma_{b} - \Gamma_{eq}\right)
.\end{equation}

Finally,  the following differential equation is solved:

\begin{equation}
\frac{\partial \Gamma_b(Z,\gamma_e)}{\partial Z} = - \frac{1}{l(Z,\gamma_e)} \left(\Gamma_b(Z,\gamma_e) - \Gamma_{eq}(Z)\right)
\end{equation}

with $\displaystyle l(Z,\gamma_e) = \frac{3 m_e c^3}{8 \pi \sigma_T} \frac{\beta_{eq}^3\Gamma_{eq}^3}{\gamma_e H}\left(1+\frac{1}{3\Gamma_{eq}^2}\right)$\\

As an example, we solve this equation and compute the actual value of $\Gamma_b$ for several values of $\gamma_e$ in two different cases: the accretion disk alone and the complete case seen section \ref{sec:gam_eq} with the accretion disk, the dusty torus, and the broad line region. In this study, we suppose $\gamma_e$ constant along the jet. Of course, this is a simplistic assumption and  a complex evolution of $\gamma_e$ along the jet can be expected in a more complete modeling, but as we will see, it has very little influence on the evolution of $\Gamma_b$ in the lowest part of the jet.
 
\begin{figure}[ht]
\begin{center}
        \includegraphics[width=\hsize]{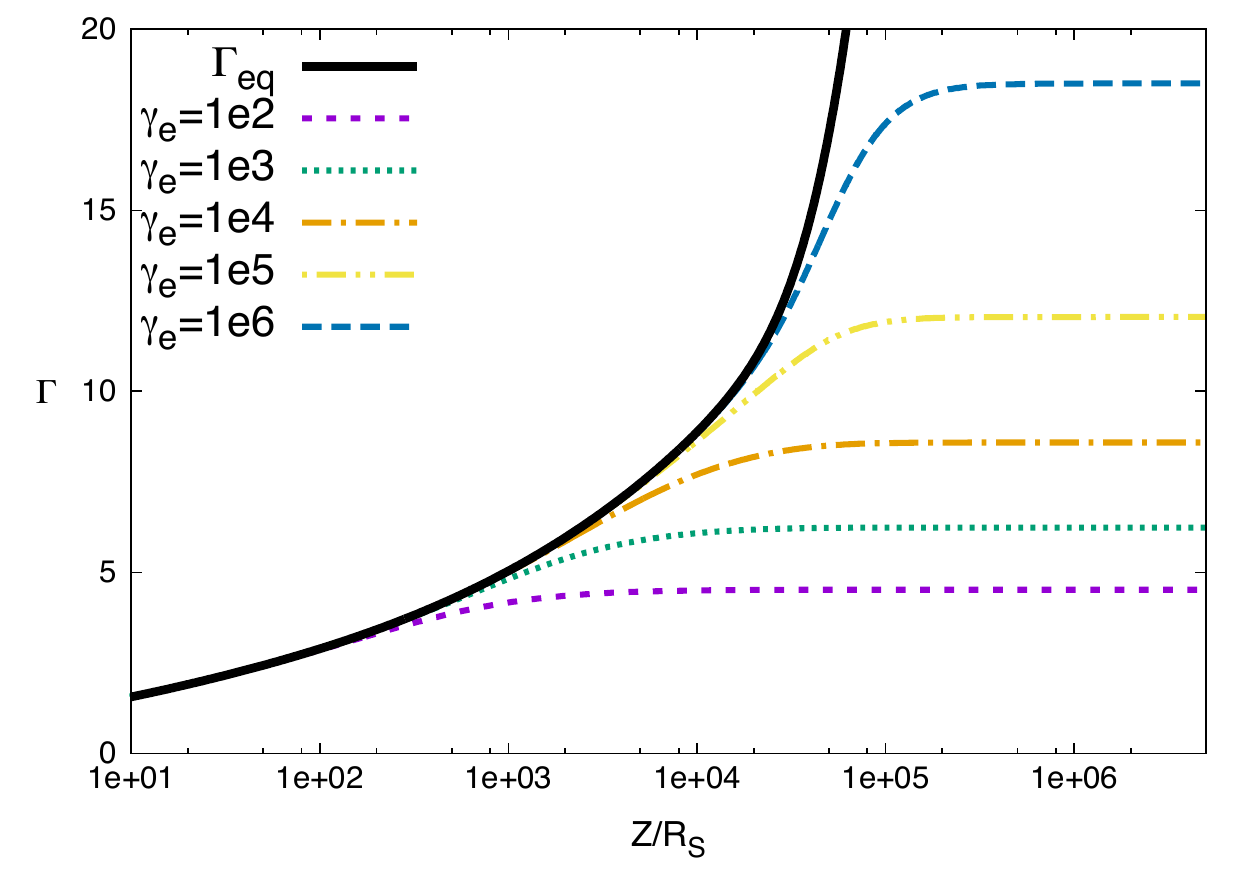}
        \includegraphics[width=\hsize]{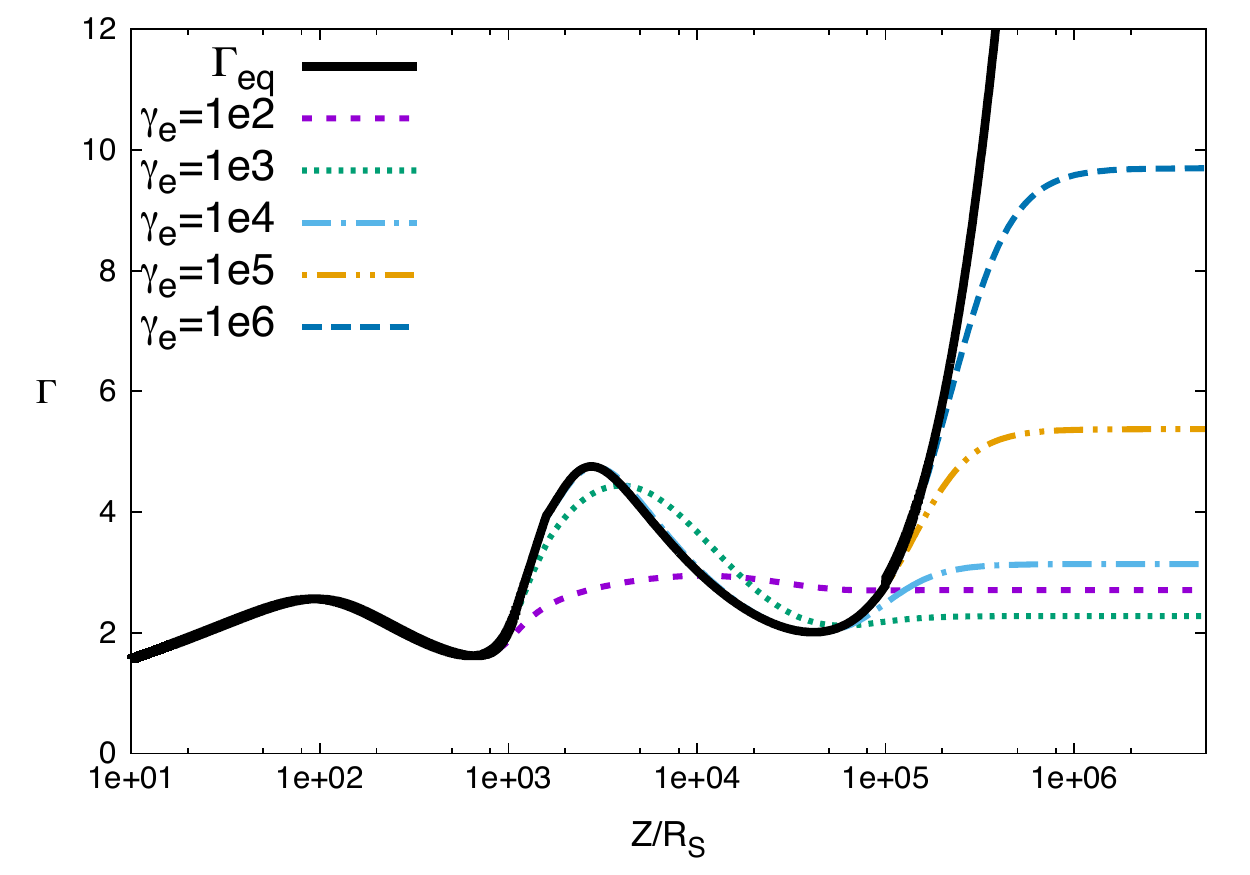}
\caption{Actual value of $\Gamma_b$ as a function of the altitude in the jet $Z/R_S$ for several values of $\gamma_e$. \textit{Top:} Standard accretion disk alone with an outer radius of $4e5 R_s$. \textit{Bottom:} Accretion disk, dusty torus, and broad line region with parameter values from table \ref{tab:sources_param}.}
\label{fig:Glim}
\end{center}
\end{figure}

Here the Compton rocket appears as a restoring force on the plasma with a stiffness constant $\displaystyle \frac{1}{l(Z,\gamma_e)}$. The relaxation length towards the equilibrium value,
$l(Z,\gamma_e)$,   is inversely proportional to $\gamma_e$ (i.e., the higher $\gamma_e$ is, the stronger the force is and the longer the plasma will actually follow $\Gamma_{eq}(Z)$) \LEt{ I had to rearrange to avoid the symbols at the beginning of the sentence. ok? Is it directly proportional? }, but  the coupling is effective in the lowest parts of the jet, no matter the value of $\gamma_e$.
At some distance (of the order of $l(Z,\gamma_e)$), the Compton rocket force slowly stops acting on the plasma. The bulk Lorentz factor\LEt{ yes? } then reaches a final Lorentz factor $\Gamma_b = \Gamma_\infty$ and follows a ballistic motion. It  can be seen that in both cases in figure \ref{fig:Glim}, high values of $\Gamma_\infty$ can be achieved with reasonable values of $\gamma_e$. The value of $\Gamma_\infty$ also depends strongly on the source geometry. Without the drag from the torus, it is easier for the jet to reach higher values of $\Gamma_\infty$. Nevertheless, even with a strong torus, the computed values of $\Gamma_\infty$ are entirely compatible with observed values by \citealt{Lister:2013gp} for the highest values of $\gamma_e$ that are totally compatible with the observed high-energy emission.

}
\section{\label{sec:emission_var}Variation in the emission}

The aim of this section is not to create a realistic model of a jet nor to explain all the variability of a single object, but to illustrate what type of variations would be induced by a variation of $\Gamma_{eq}$. We assume a jet composed of spherical emitting zones called blobs moving forward (see, e.g., \citealt{Katarzyski:2001iga}, \citealt{Boutelier:2008bga}, or \citealt{Hervet:2015wo} for \emph{blob-in-jet} models). The emergence of a blob at the base of the jet would then correspond to a flare. In our framework, this blob moves forward in the jet with an imposed bulk Lorentz factor $\Gamma_{eq}$. Because of the variation of $\Gamma_{eq}$ inducing variations of $\delta_{eq}$ for an observer, the blob emission will show some interesting changes. Therefore, a single flare will induce a complex variability as it moves along the jet, displaying associated peaks in the emission that we call echoes. To show and study these effects, we will set up a very simple jet model where we compute the synchrotron radiation (SYN), the synchrotron-self Compton (SSC), and the external Compton (EC) radiations. The following results are just examples of variations. A different model, or a complete modeling of the jet, would obviously show different results, but we can expect the variations to keep the same general features.

\subsection{Jet modeling}

To compute the emission, we first need to model the jet. The jet radius is fixed at a constant value $R_{jet} = 5 \times 10^2 R_S$. The magnetic field follows a power law $B= B_0 \left(\frac{Z}{R_S}\right)^{-1}$ with $B_0 = 6.8 \times 10^{-2} G$ for  the whole study. These values ensure that   the same global evolution is kept  between the magnetic energy density and the photon field energy density so we can compare the SCC emission and the EC emission along the jet.\\
 We have chosen a pile-up distribution for the energy distribution. It has the advantage of presenting one parameter less than a power-law distribution and concurs better with the idea of particles accelerated through the second-order Fermi process. It can be written as
\begin{equation}
n(\gamma) =  \frac{\gamma^2}{2\bar{\gamma}^3} \exp\left( - \frac{\gamma}{\bar{\gamma}} \right)
.\end{equation}

\corr{
Particles have to be energetic enough to explain the high-energy emission of AGNs,  but we also assumed  the Thomson regime to compute $\Gamma_{eq}$. The Thomson condition can be expressed as $\gamma \epsilon_s(1-\cos\theta_s)\ll 1$ in the bulk frame. Using the sources parameters given in table \ref{tab:sources_param}, the computation of $\epsilon_S(1-\cos\theta_S)$ along the jet in the rest frame of a flow at $\Gamma_{eq}$ gives a maximum value of $10^{-4}$ for the photons coming from the BLR, of $10^{-5}$ for the photons coming from the disk, and of $10^{-6}$ for the photons coming from the torus. The pile-up distribution has a mean value $< \gamma > = 3 \, \bar{\gamma}$ and drops rapidly at higher $\gamma$ because of the exponential term. 
Setting a value of $\bar{\gamma} = 10^5$ allows us to stay in the Thomson regime almost everywhere along the jet and to obtain a $\Gamma_\infty > 5$. Therefore, one can presume $\Gamma = \Gamma_{eq}$ to compute the emission at every altitude in the jet.
}

Pursuing what was said in the previous sections, each slice of the external emitting regions is a different source characterized by  four numbers: $\mu_s = \cos \theta_s$ with $\theta_s$ the incoming angle and $\dd \Omega$ the solid angle both described in figure \ref{fig:sketch} and figure \ref{fig:disk}; the temperature $T$; and the emissivity $\varepsilon$.  These parameters seen in the bulk frame depend on $\delta_b = \Gamma\left(1-\beta_b\mu_s\right)$ and are given by (parameters in the bulk rest frame are denoted by a prime) 
\begin{align}
T' & = \delta_b T &  \dd \Omega'  &= \frac{1}{\delta_b^2} \dd \Omega\\
\mu'_s & = \frac{\mu-\beta_b}{1-\beta_b \mu_s} &  \varepsilon' &=\varepsilon \notag
\end{align}

\subsection{Computation of the emission along the jet in two energy bands}

The model being set, we were able to compute broadband emission (including SYN, SSC, and EC) at every altitude along the jet for a flow at $\Gamma_b = \Gamma_{eq}$. From this, we computed the total emitted power at every altitude $\left( \frac{\dd P}{\dd \Omega} \right)_{eq}$ in two characteristic energy bands:
\begin{itemize}
\item Infrared: $[1\text{meV}-1\text{eV}]$
\item $\gamma$-rays: $[20\text{MeV}-300\text{GeV}]$
\end{itemize}

This  allows us to study the evolution of the emission as a blob of particles moves forward in the jet. 
Nevertheless, the emission $\frac{\dd P}{\dd \Omega}$ depends on $\Gamma_b$ but also on the jet model. In order to decouple the variations of emission due to the jet model from the emission due to the variations of $\Gamma_{eq}$, we computed $\displaystyle \left ( \rfrac{\dd P}{\dd \Omega} \right )_{eq}$, i.e., the emission with $\Gamma_b = \Gamma_{eq}$, and $\displaystyle \left ( \rfrac{\dd P}{\dd \Omega} \right )_{5}$, i.e., the emission with $\Gamma_b = 5$ \LEt{ ok? otherwise it looks like you computed a list of 4 things  }. This way, the function $\displaystyle \frac{\left( \dd P/\dd \Omega\right )_{eq}}{\left( \dd P/\dd \Omega\right)_5}(Z) $ is only modulated by the variations of $\Gamma_{eq}$, excluding other sources of variability.

\subsection{Time dependance of the emission in the case of a single blob moving in the jet}

In the case of a single blob traveling along the jet, we can convert the altitude into time, but owing to light time travel effects, the time between two events in the observer frame is different from the time between these two events in the lab frame. In the case of a single blob moving along the jet, the relation between $Z$ and the observation time is given by
\begin{equation}
t_{obs} = \int_{Z_{0}}^{Z_{f}} \frac{1}{\beta_{eq} c} \left( 1 - \beta_{eq} \, \mu_{obs} \right) \dd Z 
.\end{equation}

\begin{figure}[ht]
\centering
        \includegraphics[width=\hsize]{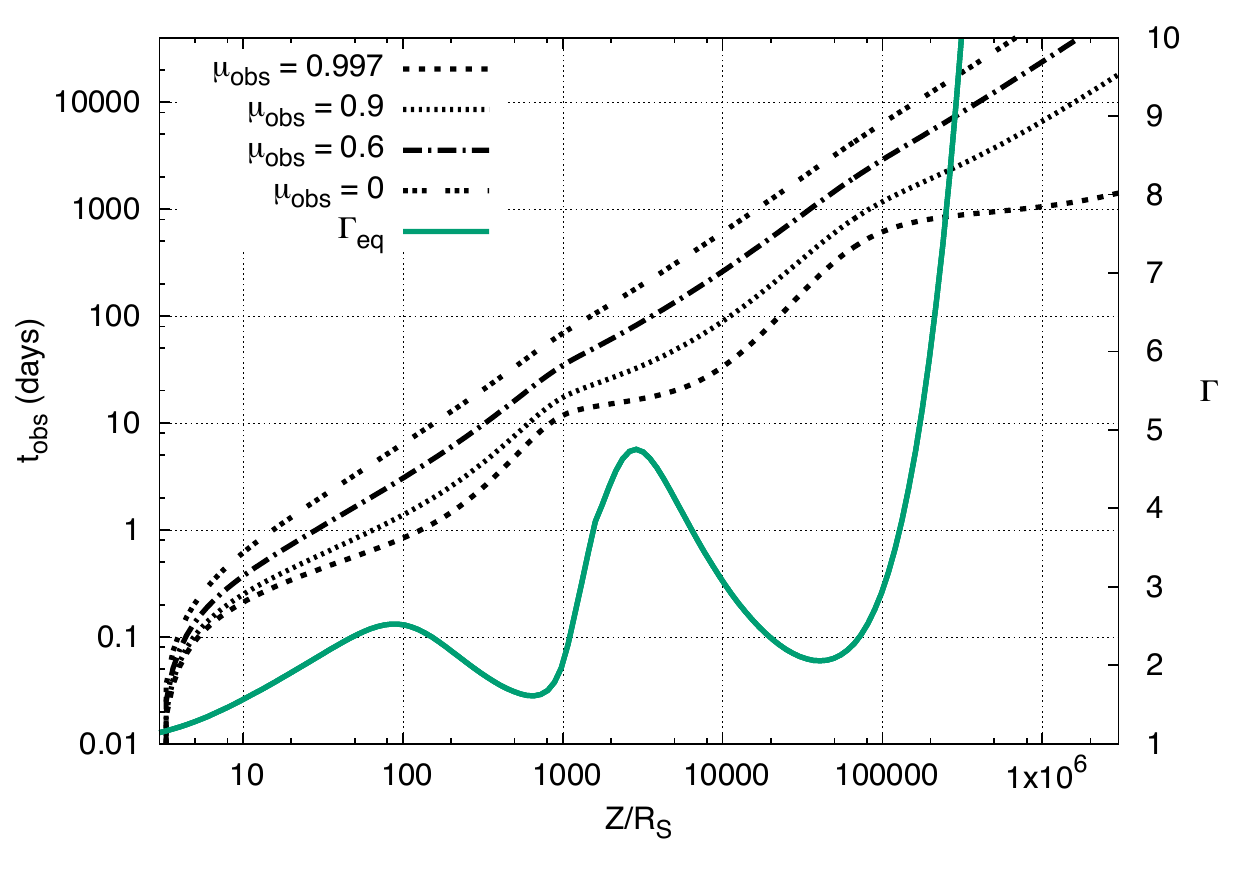}
        \caption{\label{fig:Zt} Example of evolution of the observation time as a function of the altitude in the jet for different observation angles. The solid line is the $\Gamma_{eq}$ curve, the value of $\Gamma$ is in the  scale shown on the right.}
\end{figure}

An example of the dependance of the altitude on the observed time is given in figure \ref{fig:Zt} for several observation angles. The parameters to compute $\Gamma_{eq}$ are the same as those used section \ref{sec:iobs}.

\subsection{Emission variability as a function of model parameters}
In order to compare the evolution of the emission to the evolution of $\Gamma_{eq}=$ along the jet, the study has been performed with the parameters given in table \ref{tab:sources_param}. The results are shown in figures \ref{fig:lcBLR} and \ref{fig:lctorus}.

The variations are simultaneous in infrared (IR) and in $\gamma$ rays and follow the variations of $\Gamma_{eq}$ studied in section \ref{sec:param_evol_gam}. The conclusions regarding the variations are quite similar to those on $\Gamma_{eq}$. The first echo lasting several hours is due to the acceleration of the flow followed by a deceleration due to the dragging effect from the BLR. The jet is then reaccelerated by the disk and the BLR before being dragged again by the torus, giving a second echo.

However, as the flow moves more quickly, the time contraction increases resulting in different variation timescale. 
These timescales depend on the sizes of the different sources of external emission. Because it is closer to the base of the jet, the BLR is responsible for the short timescale variations (from some hours to some days in our study). The torus, however, is responsible for variations at larger timescales (from several days to years).

\subsubsection{Influence of the BLR parameters}

\begin{figure}[ht]
\centering
\begin{minipage}[c]{1.\linewidth}
        \includegraphics[width=\hsize]{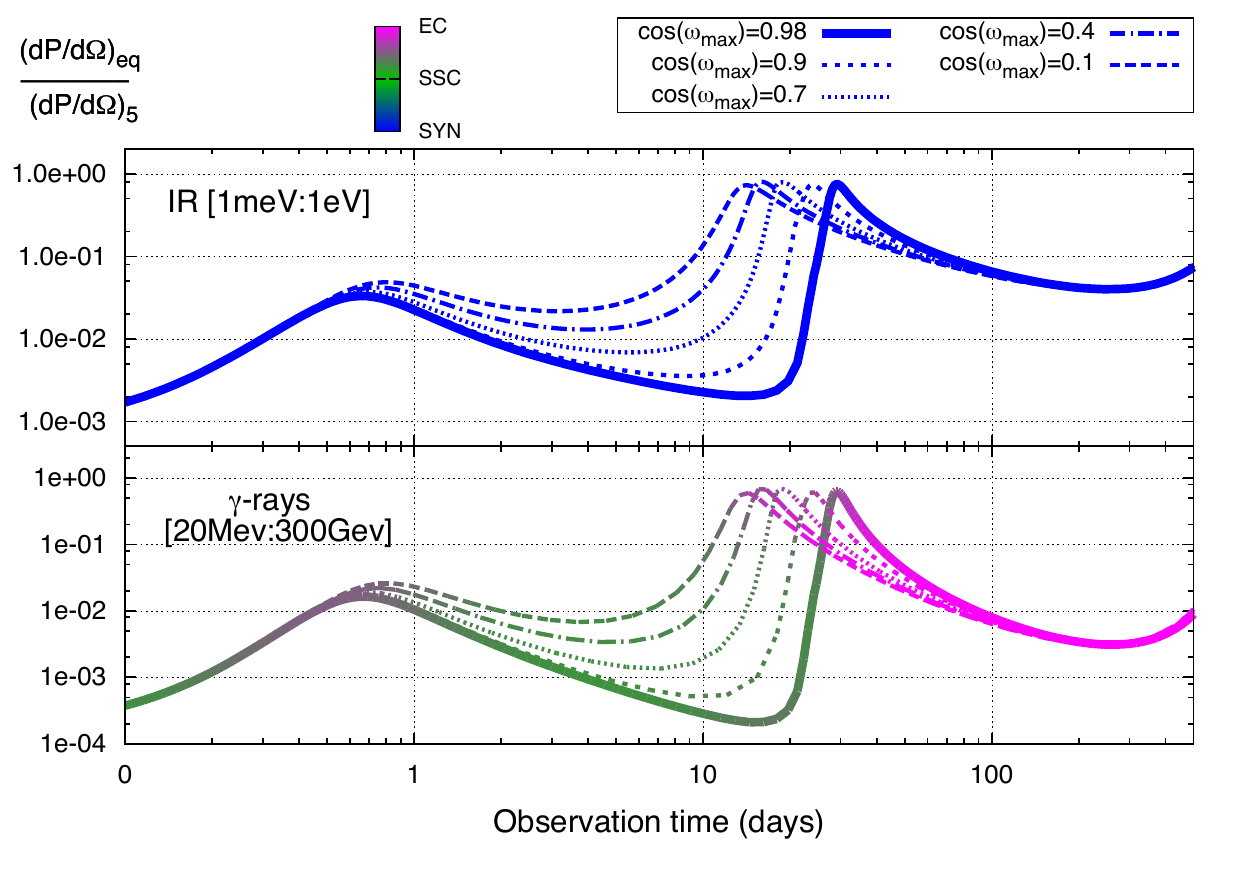}
\end{minipage} \hfill
\begin{minipage}[c]{1.\linewidth}
        \includegraphics[width=\hsize]{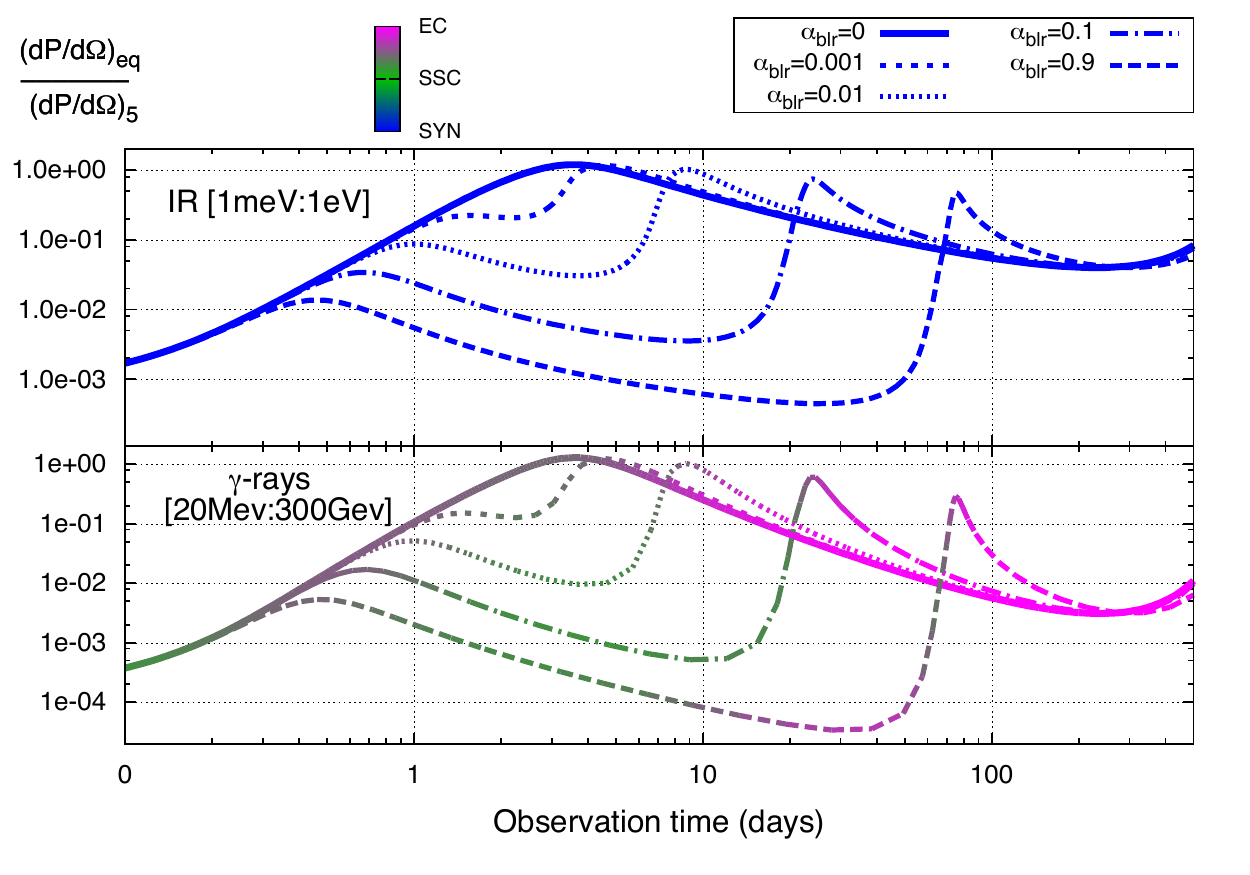}
\end{minipage}
\begin{minipage}[c]{1.\linewidth}
        \includegraphics[width=\hsize]{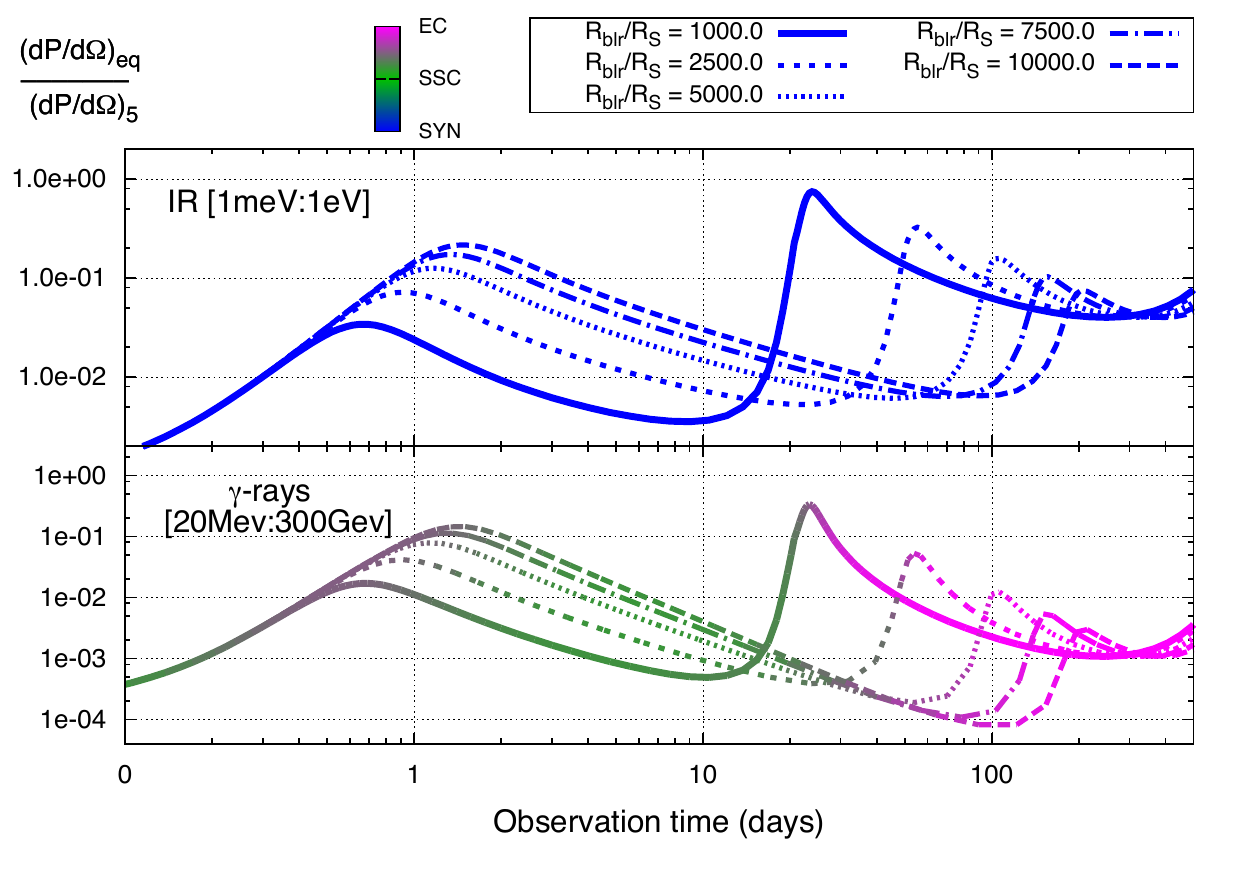}
\end{minipage}\\
\caption{\label{fig:lcBLR}Evolution of the emission in the direction of the observer $(\cos i_{obs}=0.997)$ as a  function of observed time for a range of BLR parameters. The emission of the blob following $\Gamma_{eq}$, $(\dd P/\dd\Omega)_{eq}$ is normalized by the emission of a blob at $\Gamma=5$, $(\dd P/\dd\Omega)_5$. Values of all parameters are listed in table \ref{tab:sources_param}. The color scale indicates the dominant emission process.}
\end{figure}

Figure \ref{fig:lcBLR} shows the influence of the BLR parameters on the time lag effects.
The upper plot in figure \ref{fig:lcBLR} concerns the geometrical repartition of the BLR (through its opening angle $\omega_{max}$) at constant total luminosity $L_{blr} = 0.1 L_{disk}$ \LEt{ subscript disk? }. One can see that effects are more important with a larger covering factor   because parts of the BLR closer to the jet axis have more influence on the Compton rocket.

The influence of total luminosity of the BLR (given by $\alpha_{blr}$) is more important (middle panel). For an ineffective BLR ($\alpha_{blr} = 0$), there is only one echo around day 4 \LEt{ i.e., on the fourth day? or one echo that lasts four days? I have corrected assuming the first case here } imposed by the torus,  but as the BLR becomes more important, two echoes appear, the first lasting several hours and the second  a few days. We  note that the greater the BLR, the more separated these two echoes are and the more peaked they are. Whereas the first echo  always occurs around day 1, the second  occurs later and later, from a few days to a hundred days.

Last panel concern the BLR radius. The radius has almost the same effect as $\alpha_{blr}$ as it can delay the second echo. It is also worth noting that the first and second echoes are inversely important. As the BLR grows bigger, the first echo arises later and is more important (because $\Gamma_{eq}$ is). The second echo also arises later but because it is limited by the influence of the torus, its amplitude is diminished.

Depending on the geometry and on the total luminosity of the BLR, we see that different behaviors in the time modulation of the emission are possible, which could lead to very different time variability in different objects.

\subsubsection{Influence of the torus parameters}

\begin{figure}[ht]
\centering
        \includegraphics[width=\hsize]{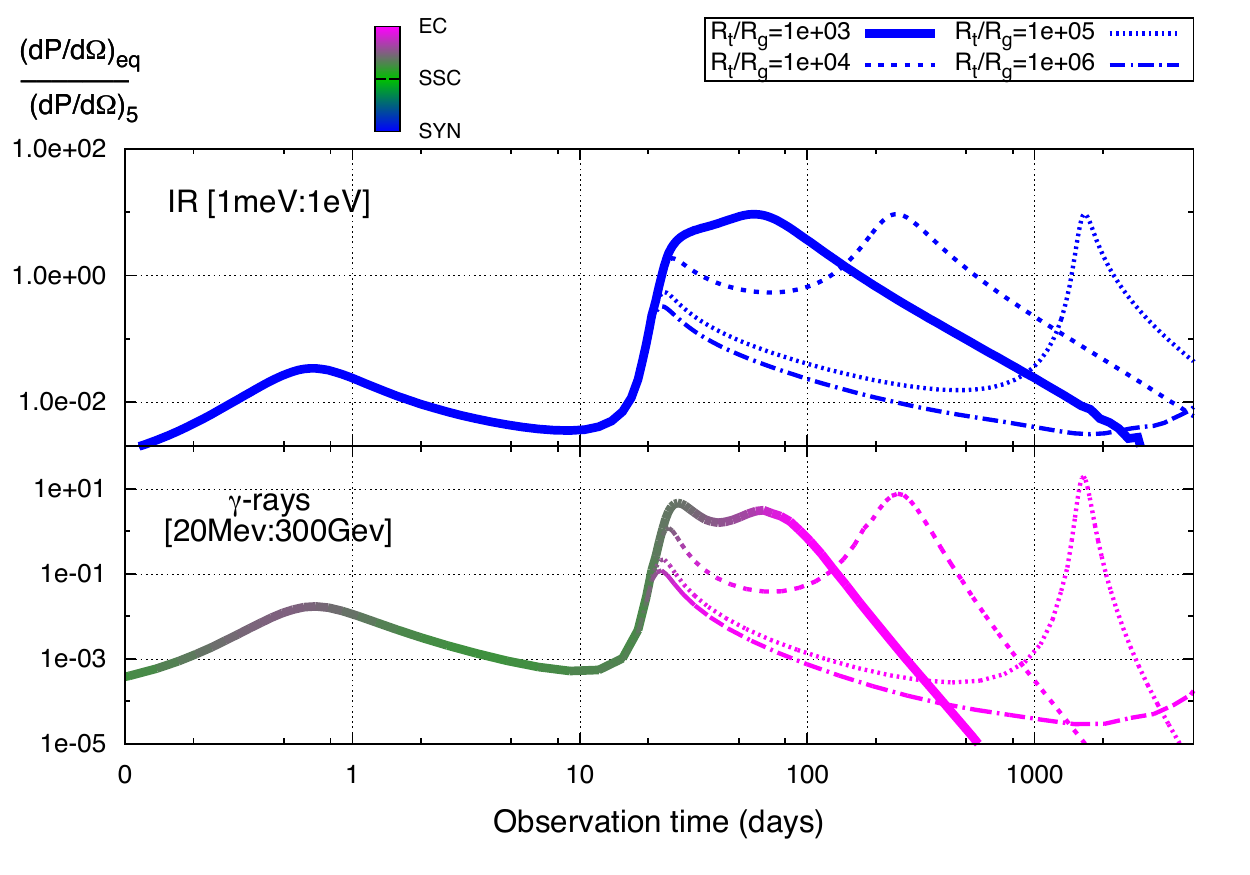}
        \caption{\label{fig:lctorus}Evolution of the emission in the direction of the observer $(\cos i_{obs}=0.997)$ as a function of observed time for a range of torus parameters. The emission of the blob following $\Gamma_{eq}$, $(\dd P/\dd\Omega)_{eq}$ is normalized by the emission of a blob at $\Gamma=5$, $(\dd P/\dd\Omega)_5$. Values of all parameters are given in table \ref{tab:sources_param}. The color scale indicates the dominant emission process.}
\end{figure}

We can see the influence of the torus size in figure \ref{fig:lctorus}. The radius of the torus evolves while we keep the continuity between the accretion disk and the torus so $D_t=r_{out} + R_t$. 

Here, we see again the two echoes imposed by the BLR and the drag from the torus after the flow crossed the BLR. As we can see, the second echo (which has  timescale of at least ten days\LEt{ yes? }), is driven by the torus size. With a greater torus size, the echo arises sooner, but is more tamed. On the contrary, a smaller torus allows the flow to reach a larger velocity, implying a stronger echo here.

At a certain point, as explained in section \ref{sec:Geq},  the flow  only accelerates, increasing the emission, but when the observer leaves the emission cone (which is highly dependent on the observation angle), the observed emission will decrease, giving the last echo. Here too, the torus size has a huge impact on the timescale of the variability. The smaller the torus, the sooner this echo arises (around day 100  here). When the torus size increases, this last echo gets delayed (up to day 2000  here), but its extent does not increase accordingly which makes it comparatively steeper. It is also interesting to note that the maximum observed value of the emission does not depend on the torus size for this kind of echo. 

\section{Conclusion}

The question of the acceleration of AGN jets is still a matter of discussion as we do not know the underlying processes or the precise speeds of the flows. The solution implied by the Compton rocket effect, viable in the two-flow paradigm, is elegant as it can naturally lead to relativistic speeds. In this work, we embrace this framework and study the influence of several external photon sources (the accretion disk, the dusty torus, and the broad line region) on the Compton rocket effect and on the induced bulk Lorentz factor. To do so, we carefully computed the resulting equilibrium bulk Lorentz factor, $\Gamma_{eq}$, of a flow driven by the Compton rocket effect taking into account the anisotropy of the emission. With several external sources, $\Gamma_{eq}$ will show important changes along the jet, leading to acceleration and deceleration phases. We studied the influence of the external sources on these patterns and the induced Doppler factor as a  function of the observation angle. We also showed that the emission of a flow following this $\Gamma_{eq}$ will experience correlated variations and that a single flare could be echoed several times. This could take part in the time-variation of these very variable objets and so we computed some examples of observed emission to illustrate our discussion. Even though these effects could not explain all the AGN variability alone, we find some interesting and non-trivial effects that could be part of the observed variability. This work could have some influence on the statistical study of AGNs and on their modeled variability. It will be followed by a more complete and more realistic model, applied to real objects to understand more precisely the role of the Compton rocket in the AGN variability.\\

\section*{Acknowledgement}
We acknowledge funding support from the French CNES agency. We also thank the referee as well as the editor for useful comments and corrections which greatly improved the manuscript.

\bibliographystyle{aa}
\bibliography{biblio}

\end{document}